\documentclass[
  aip,
  apl,
  amsmath,amssymb,
  reprint
]{revtex4-2}
\usepackage{ulem}
\usepackage{graphicx}% Include figure files
\usepackage{dcolumn}% Align table columns on decimal point
\usepackage{bm}% bold math
\usepackage[utf8]{inputenc}
\usepackage[T1]{fontenc}
\usepackage{mathptmx}
\usepackage{etoolbox}
\usepackage{braket}
\usepackage{xcolor}
\usepackage[colorlinks=true,allcolors=blue]{hyperref}

\makeatletter
\def\@email#1#2{%
 \endgroup
 \patchcmd{\titleblock@produce}
  {\frontmatter@RRAPformat}
  {\frontmatter@RRAPformat{\produce@RRAP{*#1\href{mailto:#2}{#2}}}\frontmatter@RRAPformat}
  {}{}
}%
\makeatother
\begin{document}

%\preprint{AIP/123-QED}

\title{Quantum sensing composite excitations in an anisotropic ferromagnet via a qubit}

\author{Amey S.~Rodge}
\author{Tarek Moussa}
\author{Akashdeep~Kamra}
\author{Bashab~Dey}
\email{amey.rodge@edu.rptu.de, bashab.dey@rptu.de}
\affiliation{%
Department of Physics and Research Center OPTIMAS, 
Rheinland-Pf\"{a}lzische Technische Universit\"{a}t Kaiserslautern-Landau, 
67663 Kaiserslautern, Germany
}

\date{\today}

\begin{abstract}
Ordered magnets harbor intrinsically squeezed ground states and magnonic excitations characterized by entanglement between spins and nonclassical magnon number composition. A pathway to detecting the superpositions of noneigenmode magnon number states underlying these nonclassical magnetic ground states has recently been demonstrated by utilizing a qubit coupled to the magnon mode via a {\it direct dispersive} interaction. Here, we theoretically develop this qubit spectroscopy further delineating the capabilities and limitations of this qubit spectroscopy for sensing the quantum superpositions that underlie the excited states. We demonstrate that the spectroscopy lends itself naturally to unraveling the superpositions that underlie the various quantized squeezed-magnon number states. However, excited states comprising superpositions of multiple squeezed Fock states become increasingly hard due to the large number of possible transitions, and resulting peaks, in the qubit spectroscopy thereby requiring qubits with narrower linewidths. Along the same lines, we theoretically demonstrate the qubit spectroscopy of a low amplitude coherent squeezed-magnon state analyzing the tradeoff between frequency crowding due to multiple transitions and peak linewidths. Our work lays the groundwork and design equations for deploying high-quality qubits towards sensing the composite nature of spin excitations in magnetic systems.
\end{abstract}

\maketitle

% \begin{quotation}
% The ``lead paragraph'' is encapsulated with the \LaTeX\ 
% \verb+quotation+ environment and is formatted as a single paragraph before the first section heading. 
% (The \verb+quotation+ environment reverts to its usual meaning after the first sectioning command.) 
% Note that numbered references are allowed in the lead paragraph.
% %
% The lead paragraph will only be found in an article being prepared for the journal \textit{Chaos}.
% \end{quotation}

%-----------------------------------------Intro------------------------------------------------------- %

\section{Introduction}

A solid can host different types of quasiparticles which are emergent excitations of various degrees of freedom within it and the interactions among them. Magnons are the quasiparticles which represent the quanta of collective oscillations of the spins in a magnetically ordered material\cite{Holstein1940,Yuan2022}. Since the magnons are bosonic in nature, a  magnon mode can exist in a quantum superposition of several magnon number states to form coherent states\cite{Rezende1969}, Schr\"odinger cat states\cite{Sharma2021}, squeezed states\cite{Kamra2016}, antibunched states\cite{Yuan2020_antibunch} etc.  Among them, the coherent state of magnons represents a classical excitation\cite{Glauber1963} of the magnet while the rest are nonclassical.  Although the photonic counterparts of these states have been observed and used for decades \cite{Gerry2005}, similar feat in the magnonic domain is only being witnessed recently\cite{Quirion2017,Quirion2020,WangH2021,Rani2025,Shimizu2025,Weng2026}. Significant advances are being made to probe the nonclassical magnonic states for achieving full quantum control of magnons with the prospect of integrating them with modern quantum technologies\cite{Laucht2021}and quantum computing circuits \cite{Andrianov2014,Hetenyi2022,Terhal2020,Bejarano2024}.

Squeezed states of light are of great interest as the quantum fluctuations in one of the quadratures in these states are reduced at the expense of the other in accordance with Heisenberg's uncertainty principle \cite{Walls1983,Gerry2005}.  These are nonequilibrium states of photons typically generated by driving protocols such as degenerate parametric down-conversion\cite{Slusher1985} or four-wave mixing\cite{Wu1986}. Due to suppressed fluctuations beyond the quantum limit, squeezed light has found its application in high precision LIGO measurements required for the detection of gravitational waves \cite{Abadie2011,Aasi2013}.
%The generation of squeezed photons typically relies on nonlinear driving protocols such as degenerate parametric down-conversion\cite{Slusher1985} and four-wave mixing\cite{Wu1986}.
%  %Photonic squeezed states have found their use in measurements that require high sensitivity such as the detection of gravitational waves. 
Magnons also exhibit squeezing with a graphical signature of this phenomenon being the  anisotropic quantum fluctuations of magnetization around its average value\cite{Kamra2016}. Magnon squeezing occurs in two forms -- {\it squeezed states} of magnons and {\it squeezed magnons}.
Squeezed states of magnons are nonequilibrium superpositions of magnons prepared by external means such as parametric excitation\cite{Li2019,Shimizu2025,Hioki2026}, by two-tone driving of a qubit\cite{Guo2023} or inducing Kerr nonlinearities in a cavity-magnon-qubit system\cite{Weng2026} and   their signatures have been observed by time-resolved measurements of the magnetic fluctuations and Wigner tomography \cite{Shimizu2025,Weng2026,Hioki2026}. On the other hand, squeezed magnons are the natural quasiparticle excitations of an anisotropic ferromagnet whose squeezing can be attributed to weak spin non-conserving interactions within the system and energy minimization\cite{Kamra2016,Kamra2016_hybrid,Kamra2020}. Unlike the case of squeezed state of the magnons discussed earlier, here the squeezing arises at equilibrium due to the intrinsic properties of the magnet. %Squeezing in the ground state rewards it with additional stability against decay, making a ferromagnet an attractive avenue to investigate magnon-squeezing. % This feature is absent for a photon mode in which the squeezing has to be induced externally.
%This makes a ferromagnet an attractive platform to investigate a desired quantum state of squeezed magnons.
A squeezed magnon is a composite excitation bearing an internal quantum superposition of odd magnon number states \cite{Kamra2020,Roemling2023}. While the shape of the quantum fluctuations is a graphical way to discuss the squeezing\cite{Shimizu2025,Hioki2026}, the quantum superpositions underlie all the effects such as entanglement\cite{Zou2020,Elyasi2020},  spin current shot noise \cite{Kamra2016} etc. and can provide the distinct  fingerprints required to identify these states \cite{Roemling2023}.

Quantum sensing using superconducting \cite{Degen2017,PatelDesai2025,Fink2024,Kakuyanagi2023,DanilinWeides2021} and spinful defect qubits \cite{Casola2018,Xu2023,Simon2022, Sar2015,Page_2019,Dolgirev2022,Chatterjee2022,Bhattacharyya2024,Melendez2025,Machado2023} have gained immense impetus in recent times due to their ability to detect plethora of solid state phenomena. A superconducting qubit has been used as a probe to sense the superpositions of number states of a bosonic mode \cite{Schuster2007,Arrangoiz-Arriola2019} through boson number-dependent shifts in the qubit's frequency\cite{Faria1999,Gambetta2006}.  %\sout{This sensing protocol works in the dispersive regime of  a coherently coupled bosonic mode-qubit system when boson number-dependent shifts occur in the qubit's frequency, observed through splitting of its spectroscopic peaks\cite{Faria1999,Gambetta2006}. The heights of the peaks are directly related to the corresponding boson number distribution.} 
This principle has also been applied to sense the nonequilibrium superpositions of number states of the  eigenmode magnon of a ferromagnet \cite{Quirion2017}, including its single magnon state\cite{Wang2023,Quirion2020}. However, this protocol does not reveal the %\sout{quantum nature}
internal composition of the eigenmode  magnon itself. An eigenmode magnon is a true composite excitation of the magnet which intrinsically hosts quantum superpositions of noneigenmode magnon number states that represent delocalized spin-flips. Squeezed magnons in an anisotropic ferromagnet\cite{Kamra2016}, two-mode squeezed magnons in an antiferromagnet\cite{Kamra2019} and fractionalized excitations in quantum spin liquids\cite{Savary2017,Zhou2017,Broholm2020} are some prominent examples of composite spin excitations.   % %The magnon number statistics is revealed by performing the qubit spectroscopy. 

%A direct dispersive interaction between a spin qubit and the noneigenmode magnon {\color{magenta} can be used to 
 It has been theoretically demonstrated that a qubit can be used as a probe to resolve the intrinsic magnonic superpositions of the ground state in an anisotropic ferromagnet\cite{Roemling2023} as well as an antiferromagnet~\cite{Roemling2024,Roemling2025}. The idea is to engineer a dispersive interaction between the noneigenmode magnon and a qubit via which the information about the intrinsic magnonic superpositions gets imprinted in the qubit frequency.  This provides direct evidence of the squeezing in the ground state of the magnet i.e. the vacuum of the composite excitations. This accomplishment raises the question:  Can we resolve the internal superpositions which underlie a single composite excitation or a coherent state of such composite excitations? Addressing this question can be a crucial step forward in extracting the quantum information contained in an actual quasiparticle excitation of a system, and may also provide a path towards detecting elusive quasiparticles such as the fractionalized excitations of a quantum spin liquid\cite{Savary2017,Zhou2017,Broholm2020}, for example.

%\sout{Since we already have a recipe to detect the magnon squeezing at equilibrium, it is desirable to see whether this recipe can be used to decipher the quantum composition of any squeezed magnon number state, which is an eigen-excitation of the ferromagnet.} 
In this work, we demonstrate that a qubit can be used as a probe to %\sout{spectroscopy can indeed} 
resolve the quantum superpositions that make up the squeezed magnon excited states and also their arbitrary superpositions. The results are supported by numerical simulations of the qubit spectroscopy. %\sout{We delineate the analytical forms of the superpositions which constitute some of the squeezed magnon excited states. We derive the expressions of the multivalued qubit frequencies which contain information about the magnonic statistics of each of these states.} 
 In order to model an experimental implementation of the qubit spectroscopy, we also perform a simulation of the sensing protocol considering all the decay channels of the combined magnon-qubit system weakly coupled to a thermal bath. We demonstrate the generation of a coherent state of squeezed magnons and its subsequent sensing using a qubit as the probe.  We show that spectral crowding and finite temperature effects can pose significant challenges to the sensing. 
 
 The article is organized as follows. In Sec. \ref{Model}, we describe the theoretical model of the system. %In Sec. \ref{anisotropic-magnet}, we discuss the quantum model of an anisotropic ferromagnet, which forms the basis of our problem. In Sec. \ref{squeezed-magnon}, we explain the emergence of squeezed magnons as the eigen-excitations of the system. In Sec. \ref{superpositions}, we discuss the distinctive nature of quantum superpositions that define the squeezed magnon states. In Sec. \ref{coupled-system}, we discuss the dispersively coupled magnon-qubit system and show how it results in qubit-dependent squeezing. 
In Sec. \ref{Spectroscopy}, we elucidate how the qubit spectroscopy can sense the internal superpositions in a squeezed magnon %In Sec. \ref{simulation}, we discuss a numerical protocol to simulate the corresponding spectroscopy. In Sec. \ref{Simulation-results}
and present the results of our simulation. In Sec. \ref{detection}, we demonstrate the generation and sensing of a nonequilibrium state of squeezed magnons. In Sec. \ref{limitations}, we highlight the limitations of the spectroscopic protocol. In Sec. \ref{conclusion}, we summarize our results.

%------------------------------------------------------------------------------------------------ %

%-----------------------------------------Section 2------------------------------------------------------- %
\section{Theoretical model}\label{Model}

In this section, we provide a qualitative discussion of the key phenomena at play and introduce the general mathematical framework for the problem.
%\subsection{Anisotropic ferromagnet}\label{anisotropic-magnet}
We consider an anisotropic ferromagnet described by the spin Hamiltonian $(\hbar=1)$\cite{Skogvoll2021}
\begin{equation}
\begin{aligned}
\hat{H}_{\text{fm}} ={}&
- \frac{J}{2} \sum_{\langle i,j\rangle} 
\hat{\bm S}_{i} \cdot \hat{\bm S}_{j}
+ |\gamma| \sum_{i} 
\left( \hat{\bm S}_{i} \cdot \mu_{0} \hat{\bm H}_{\text{ext}} \right) \\
&+ \sum_{i}\left(
K_{x} \hat{S}_{ix}^{2}
+ K_{y} \hat{S}_{iy}^{2}
+ K_{z} \hat{S}_{iz}^{2}
\right).
\end{aligned}
\label{eq:FerromagneticHamiltonian}
\end{equation}
The first term in Eq.~(\ref{eq:FerromagneticHamiltonian}) describes the exchange interaction between neighboring spins which is responsible for the ferromagnetic order, the second term denotes the Zeeman interaction of the spins with an external magnetic field $\hat{\bm H}_{\mathrm{ext}}=-H_{0}\hat{z}$ which tries to align the spins along $\hat{z}$ direction and the third term represents an interaction which encapsulates magnetocrystalline and shape anisotropies in the system. Here, $J>0$ denotes the exchange energy, $\gamma=-|\gamma|$ the gyromagnetic ratio and $\mu_{0}$ the magnetic permeability. The indices $i$ and $j$ label the spin sites in the crystal, $\hat{\bm S}_{i(j)}$ is the spin operator at site $i(j)$ and $\langle i,j\rangle$ implies that the summation is only over the nearest neighbor pairs $(i,j)$. 

The strengths of the magnetic anisotropies along different directions are characterized by the coefficients $K_{x}$, $K_{y}$, and $K_{z}$. In absence of anisotropy in the plane perpendicular to the applied field i.e. $K_x,K_y=0$, the quantum mechanical ground state resembles the classical state with all spins pointing along $\hat{z}$ i.e.  $|\uparrow\uparrow\uparrow\uparrow\uparrow\uparrow\dots\rangle$ and the low energy excitations are the bosonic quasiparticles called magnons, representing spin flips delocalized throughout the material\cite{Kittel1953}.  For $K_x, K_y\neq0$ and $K_x\neq K_y$, $|\uparrow\uparrow\uparrow\uparrow\uparrow\uparrow\dots\rangle$ is no longer the quantum ground state and the corresponding excitations are different from the `spin-flip' magnons. %The new ground state is a $squeezed$ vacuum and the new eigen-excitations are called $squeezed$ magnons. 
%\sout{The objective of our work is to formulate a protocol for detecting the new quantum excitations of the system.}

%\subsection{Squeezed magnons}\label{squeezed-magnon}
The new ground state and low energy eigen-excitations of the system can be conveniently described by transforming $\hat{H}_\text{fm}$ to the magnon Hamiltonian through the Holstein-Primakoff transformations\cite{Holstein1940}. Retaining only the uniform (${\bf k}=0$) noneigenmode magnon\cite{Skogvoll2021} (assuming a small magnet) and denoting its annihilation and creation operators as  $a$ and $a^\dagger$ respectively, the magnon Hamiltonian is expressed as ~\cite{Roemling2023}
\begin{equation}
    \hat{{H}}_a = A \hat{a}^{\dagger} \hat{a} + B \hat{a}^2+ B^{*} \hat{a}^{\dagger 2}
\label{eq:FM_Hamiltonian_k0}
\end{equation}
where $A = |\gamma| \mu_{0} H_{0} + (K_{x} + K_{y} - 2K_{z})S$ and $B = S(K_{x} - K_{y})/2$.  Here, $S$ is the spin length and the parameter $B$ characterizes the relative anisotropy in the  plane ($x$-$y$) perpendicular to the  applied magnetic field. %For equal anisotropy along $x$ and $y$ directions (i.e. $B=0$), the low energy excitations of the ferromagnet are the Fock states $\{|n\rangle\}$ of the magnon number operator $\hat{a}^\dagger \hat{a}$. %parameterizing the anisotropy\cite{PhysRevApplied.16.064008} in $x$-$y$ plane. 
  When the anisotropy along one direction exceeds the one in the other (i.e. $B\neq0$), $\hat{{H}}_{a}$ is not diagonal in the Fock basis. It can be diagonalized via the Bogoliubov transformation $\hat{\alpha} = \hat{a}\cosh{r} + \hat{a}^{\dagger} e^{i\theta} \sinh{r}$ with 
\begin{equation}\label{rtheta}
    r = \frac{1}{2} \text{arctanh} \left(\dfrac{2|B|}{A}\right)~~~\text{and}~~~~ e^{i\theta} =\frac{B^*}{|B|},
\end{equation}
resulting in a new bosonic Hamiltonian $ H_\alpha=\omega_{\alpha}\hat{\alpha}^{\dagger}\hat{\alpha}$. Now, $
    \omega_{\alpha} = \sqrt{A^{2} - 4|B|^2}
$
is the magnon eigen-frequency which is a function of the relative anisotropy and $\{|n\rangle_\alpha\}$, the number states of $\alpha^\dagger \alpha$, constitute the eigen-excitations of the system.  These are called $squeezed$ magnons as the quantum fluctuations in the spin quadratures $S_x$ and $S_y$  are not symmetric  in these states but squeezed along either $x$ or $y$ direction\cite{Kamra2020}. The parameter $r$ in Eq. (\ref{rtheta}) is a measure of the degree of squeezing. 
 %It is important to note that $\hat{\alpha}$ is not just a linear combination of bosonic creation ($\hat{a}$) and annihilation ($\hat{a}^{\dagger}$) operators, but it results from a transformation characteristic of the squeezing phenomenon applied to the original bosonic mode \cite{gerry2023introductory,scully1997quantum}. It is related to the magnon $\hat{a}$ via a single-mode squeeze operator $\hat{{S}}(\xi) = \exp \left({\frac{\xi ^{*}}{2} \hat{a}^{2} -\frac{\xi}{2} \hat{a}^{\dagger 2}} \right)$, where $\xi=re^{i\theta}$ is a complex squeezing factor.
  %\textcolor{red}{Note that in contrast, %The presence of anisotropy in the Hamiltonian $\hat{H}_{\mathrm{FM}}$ causes the eigenstates to be squeezed~\cite{gerry2023introductory,PhysRevApplied.16.064008,Kamra_2020}, such that the bare magnon operators $\hat{a}$ no longer diagonalize the system. 
  The squeezed magnon states are related to the Fock magnon states through a unitary transformation given by the squeeze operator\cite{Satyanarayana1985,Kim1989} $\hat{{S}}(\xi) = \exp \left({\frac{\xi ^{*}}{2} \hat{a}^{2} -\frac{\xi}{2} \hat{a}^{\dagger 2}} \right)$ such that
\begin{equation}\label{squeeze}
    \ket{n}_{\alpha} = \hat{{S}}(\xi)\ket{n}
    =\sum_{m}C_{m,n}(\xi) \ket{m},
\end{equation}
where $\xi=re^{i\theta}$ is the complex squeezing factor, and the coefficient $C_{m,n}(\xi) = \bra{m}\hat{S}(\xi)\ket{n}$ denotes the $(m,n)$ matrix element of the squeeze operator in the Fock basis. Hence, the new composite excitations of the system are formed by a quantum superposition of several Fock magnon number states.   %Even the squeezed magnon vacuum $\ket{0}_\alpha$, which represents the new ground state of the system, harbors these superpositions. 

%Anisotropy in the $x$-$y$ plane therefore leads to squeezed eigenstates of the ferromagnet~\cite{gerry2023introductory}. These constitute the true eigenstates of the system and resolving them is the primary objective of this work.

 \subsection{Quantum superpositions in squeezed magnons}\label{superpositions}

% Using Eq.(\ref{eq:eff_squeez}), this fact translates to \begin{equation}
%\ket{n}_g
 %   = \sum_{m=0,1,2...} {C}_{2m,n}(\xi_{\mathrm{eff}})\ket{2m}_e~~~\text{for even}~n 
%\end{equation}
%and
%\begin{equation}
 %   \ket{n}_g
  %  = \sum_{m=0,1,2...} {C}_{2m+1,n}(\xi_{\mathrm{eff}})\ket{2m+1}_e ~~~\text{for odd}~n 
%\end{equation}
%between the squeezed states.
Since the argument of the exponential in the squeeze operator creates and annihilates two magnons, a squeezed magnon number state $\ket{n}_\alpha$ with even (odd) $n$ is a superposition of Fock magnon states $\{\ket{m^\prime}\}$ with even (odd) values of $m^\prime$.
The analytical form of superpositions constituting  the squeezed‑vacuum state $\ket{0}_\alpha$ can be expressed as
$\ket{0}_{\alpha} = \sum_{m=0}^{\infty}C_{2m,0}(\xi) \ket{2m}$ with expansion coefficients\cite{Gerry2005,Walls2025QuantumOptics}
\begin{equation}
    {C}_{2m,0}(\xi) = \dfrac{(-1)^{m}}{\sqrt{\cosh(r)}} \left[e^{i\theta} \tanh(r)\right]^{m} \dfrac{\sqrt{(2m)!}}{2^{m} m!}.
    \label{coeff0}
\end{equation}
 Although the analytical expressions of the coefficients $C_{m,n}(\xi)$ are available in the literature\cite{Satyanarayana1985,Kim1989}, we provide a systematic derivation of the closed‑form expressions for superpositions contained in the squeezed magnon single ($\ket{1}_{\alpha}$) and double excitations ($\ket{2}_\alpha$) in the Supplementary Material (SM) [\ref{SM}].  One can also obtain the higher excitations of the squeezed magnon states in a similar manner. %The expressions reveal the detailed composition of each state and enable a direct comparison between magnonic superpositions resolved using the qubit and the exact analytical results.
 The magnon basis-expansion of a squeezed-magnon single-excitation is derived as $\ket{1}_\alpha =\sum_{m=0}^{\infty} C_{2m+1,1}(\xi)\ket{2m+1}$
with expansion coefficients
\begin{equation}\label{coeff1}
    C_{2m+1,1}(\xi) =
    \frac{(-1)^m}{\cosh^{3/2}(r)}
    \frac{\sqrt{(2m+1)!}}{2^m m!}
    \left[e^{i\theta}\tanh(r)\right]^m.
\end{equation}
 Similarly, the squeezed-magnon double excitation is obtained as 
\begin{equation}
    \begin{aligned}
    \ket{2}_{\alpha} &= \sum_{m=0}^{\infty}{G}_{2m +2,2}(\xi) \ket{2m +2}+ \sum_{m=0}^{\infty}{D}_{2m,2}(\xi) \ket{2m}
    \end{aligned}
    \label{ket2_g}
\end{equation}
where the coefficients are 
\begin{equation}\label{coeff2}
    \begin{aligned}
    {G}_{2m +2,2}(\xi) &=\left(\dfrac{\sqrt{(2m+2)!}}{2^{m} m!}\right) \dfrac{\left[-e^{i\theta} \tanh(r)\right]^{m} }{\sqrt{2\cosh(r)}}, \\
    {D}_{2m,2}(\xi) &= \left((2m+1) \dfrac{\sqrt{(2m)!}}{2^{m} m!}\right) \dfrac{(-1)^{m}\left[e^{i\theta} \tanh(r)\right]^{m+1}}{\sqrt{2\cosh(r)}} .
    \end{aligned}
\end{equation} 
 %This feature will be observed during the resolution of the magnonic superpositions through qubit spectroscopy.

Resolving the superpositions of any squeezed magnon state $\ket{n}_\alpha$ is one of the keys to detecting these states.
%It has been theoretically demonstrated that a qubit can be used as a probe to resolve the superpositions in the ground state of the system i.e. the squeezed magnon vacuum~\cite{Roemling2023}. The idea is to engineer a dispersive interaction between the magnon mode and a qubit via which the information about the magnonic superpositions gets imprinted in the qubit frequency.  %so that quantum information contained in the squeezed magnon vacuum is imprinted onto the qubit frequency. 
%This was realized via a direct dispersive interaction between the qubit and the magnon mode. 
%Consequently, the qubit spectroscopy reveals the equilibrium magnonic superpositions.
%This is fundamentally different from the case where a nonequilibrium superpositions of magnons, created by a coherent drive, could be resolved via qubit spectroscopy. 
%We extend this approach to probe the superpositions contained in any excited state of the system.  %We also use this protocol to detect the superpositions of different squeezed magnon states. We derive analytical expressions of the multivalued qubit frequencies which
%serve as signatures of the occupied magnon number states. We provide exact analytical expressions of the magnonic superpositions comprising the squeezed states and generalize the concept of ``effective squeezing,'' which plays a key role in the underlying physics. We further discuss the limitations and challenges associated with resolving the \textcolor{red}{coherent excitations}. 

\subsection{Dispersive interaction between the magnon mode and a qubit}
\label{coupled-system}
An effective dispersive coupling $\sim \hat{m}^\dagger \hat{m} \sigma_z$ has been engineered between an eigenmode magnon (annihilation operator: $\hat{m}$) and a superconducting qubit (operator $\sigma_z$)\cite{Quirion2017}. This interaction makes the qubit frequency a function of the eigenmode magnon number. As a result, when a nonequilibrium superposition of the number states is generated, the qubit spectrum peaks at multiple frequencies corresponding to the occupied number states, the height of a peak being proportional to the occupation probability of the number state. This allows the qubit to resolve number states of the eigenmode magnon.

Interfacial interaction between the spins of an anisotropic ferromagnet and a semiconductor spin qubit \cite{Skogvoll2021} can also give rise to a direct dispersive interaction $\sim a^\dagger a\sigma_z$  between a  noneigenmode magnon (annihilation operator: $\hat{a}$) of the ferromagnet and the qubit. The coupling with the noneigenmode magnon instead of the eigenmode (annihilation operator: $\hat{\alpha}$) offers the qubit a fundamentally new sensing ability -- it can now resolve the internal superpositions of the $\hat{a}$ mode states contained in  $\hat{\alpha}$ mode\cite{Roemling2023}. %\Tarekcom{I don't find it clear what we are using for our model and where our dispersive coupling is coming from}
  The Hamiltonian describing the squeezed magnon-qubit system  with a dispersive  coupling through the noneigenmode magnon is given by \cite{Roemling2023}
\begin{equation}
    \hat{{H}}_{\text{sys}} =  A\hat{a}^{\dagger} \hat{a} + B \hat{a}^2+ B^{*} \hat{a}^{\dagger 2} + \dfrac{\omega_{q}}{2} \hat{\sigma}_{z} + \chi \hat{a}^{\dagger} \hat{a} \hat{\sigma}_{z},
    \label{System_Hamiltonian}
\end{equation}
where $\omega_q$ is the bare qubit frequency, $\chi$ is the dispersive strength and $\sigma_z=\ket{e}\bra{e}-\ket{g}\bra{g}$ is the qubit operator with $\ket{e}$ (excited) and $\ket{g}$ (ground) constituting its basis states. 
Projecting the Hamiltonian onto the qubit basis yields
\begin{equation}
    \hat{H}_{\mathrm{sys}}
    =
    \begin{pmatrix}
        \hat{H}_e & 0 \\
        0 & \hat{H}_g
    \end{pmatrix}
    \equiv
    \begin{pmatrix}
        \bra{e}\hat{H}_{\mathrm{sys}}\ket{e} & 0 \\
        0 & \bra{g}\hat{H}_{\mathrm{sys}}\ket{g}
    \end{pmatrix},
    \label{eq:system_hamiltonian_matrix}
\end{equation}
where $H_{g(e)}= [A-(+)\chi]~\hat{a}^{\dagger} \hat{a} + B \hat{a}^2+ B^{*} \hat{a}^{\dagger 2} -(+) \dfrac{\omega_{q}}{2}$. This implies that the dispersive coupling results in different squeezings $r_g$ and $r_e$ in the ground and excited state of the qubit, respectively, where $r_{g(e)}$ is obtained by transforming $A\to A -(+)\chi$ in Eq. (\ref{rtheta}).  Rewriting the Hamiltonian in the squeezed magnon basis $\hat{\alpha}_{g(e)}=\hat{a}\cosh{r}_{g(e)} + \hat{a}^{\dagger} e^{i\theta} \sinh{r}_{g(e)}$ of the ground (excited) state of the qubit, we get \begin{equation}
    \hat{H}_{g(e)} = \omega_\alpha^{g(e)}\hat{\alpha}_{g(e)}^\dagger\hat{\alpha}_{g(e)} -(+)\dfrac{\omega_{q}}{2}+\dfrac{\omega_\alpha^{g(e)}-[A-(+)\chi]}{2}
\end{equation} 
where
$\omega_\alpha^{g(e)}=\sqrt{[A-(+)\chi]^{2} - 4|B|^2}$ are eigen-frequencies of the squeezed magnons, which now depend on the state of the qubit. %This shows that the magnon mode has different squeezings in the ground and excited states of the qubit. 
We will demonstrate shortly how the difference in the squeezings allows us to detect the magnonic superpositions present in these states.

\begin{figure*}
    \centering
    \includegraphics[width=0.85\textwidth]{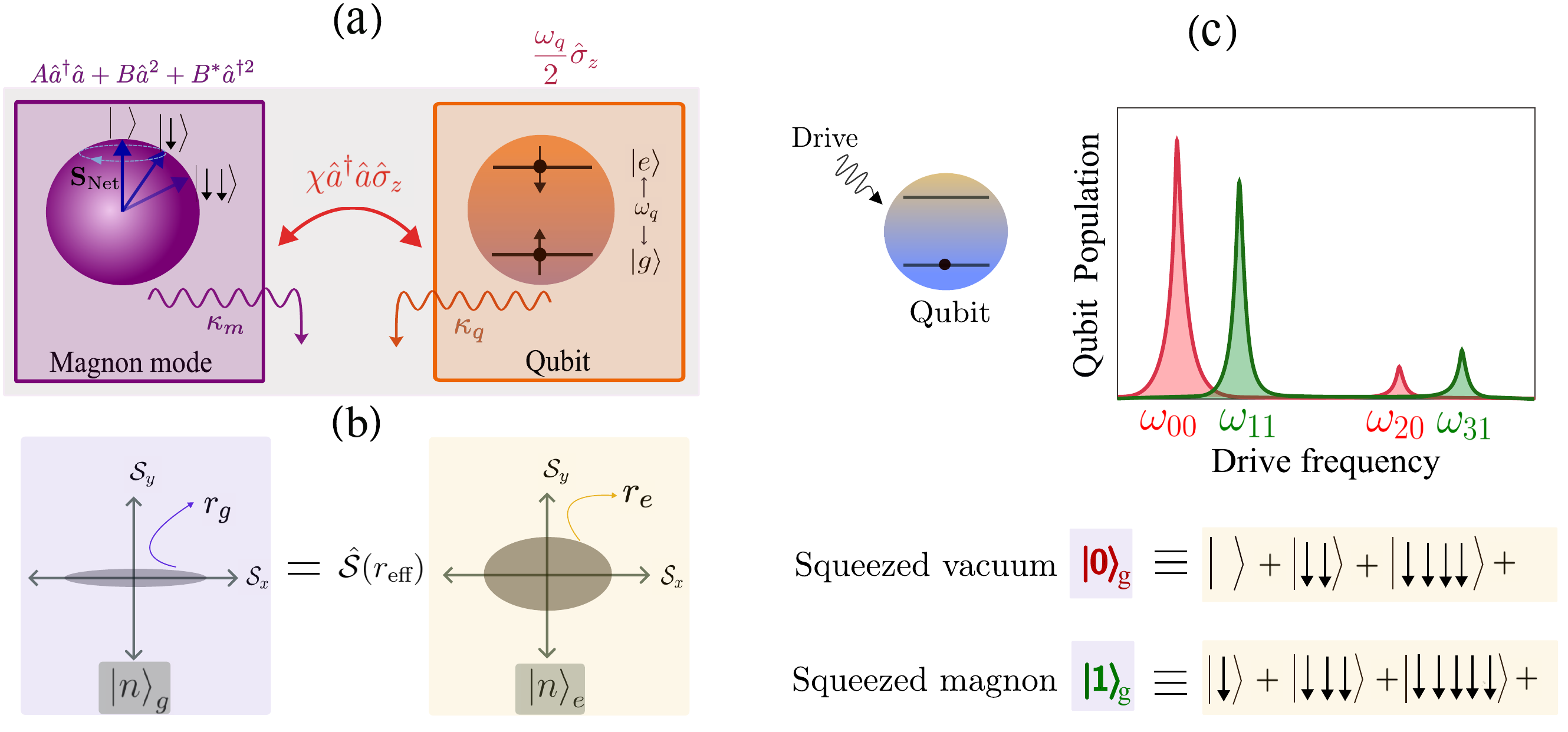}
    \caption{ 
    (a) Schematic depiction of the system. The noneigenmode magnon (left) with annihilation operator $\hat a$, having a term $B\hat a^2+ B^* \hat a^{\dagger2}$ which gives rise to intrinsic squeezing,  is coupled to a spin qubit (right) via a direct
    dispersive interaction $\chi \hat a^\dagger a\hat\sigma_z$. The magnon and qubit decays are captured through their decay rates $\kappa_m$ and $\kappa_q$, respectively.
    (b) %The ground-state squeezed-magnon $\ket{n}_{g}$ is squeezed relative to the excited-state squeezed-magnon $\ket{n}_{e}$ with an effective squeezing $r_{\mathrm{eff}} = r_{g} - r_{e}$.
    Qubit-dependent squeezing.  The shape of the quadrature ellipses for the squeezed magnon number states
    in the ground ($\ket{n}_{g}$) and excited $(\ket{n}_{e}$) states of the qubit respectively show different degrees of squeezing as
    $r_{g}>r_{e}$. The state $\ket{n}_{g}$ state is  effectively squeezed with respect to the $\ket{n}_{e}$  state, as shown by the application of the squeeze operator $\hat{S}(r_\text{eff})$.
    (c) Qubit spectroscopy.  %for the squeezed-magnon vacuum and the single-excitation state. 
    The qubit is driven coherently and its steady-state  population is plotted as a function of the drive frequency $\omega_d$. When the system is in the squeezed magnon vacuum $|0\rangle_g$, it has internal superpositions of even magnon number states $\ket{0}, \ket{2}, \ket{4}, \dots$ states due to which the qubit spectrum peaks at $\omega_d = \omega_{00}, \omega_{20}, \omega_{40}, \dots$ (marked red)\cite{Roemling2023}.  Similarly, when the system is a squeezed magnon single excitation state  $\ket{1}_g$, the spectrum peaks at $\omega_d = \omega_{11}, \omega_{31}, \omega_{51}, \dots$ (marked green) as $\ket{1}_g$ contains superpositions of odd magnon number states $\ket{1}, \ket{3}, \ket{5}, \dots$. The heights of the peaks carry the information of the probability of occupation of the number states, thereby allowing us to resolve the superpositions.}   %showing distinct resonance peaks associated with transitions from the squeezed-magnon vacuum $\ket{0}_{g}$ (red) and the single-excitation state $\ket{1}_{g}$ (green). The peak positions $\omega_{00}$, $\omega_{11}$, $\omega_{20}$,and $\omega_{31}$ encode the quantum superpositions underlying the squeezed-magnon states.}
    \label{fig:system}
\end{figure*}

A schematic of the hybrid qubit--magnon system is shown in Fig.~\ref{fig:system}(a). For the coupled system, we denote the squeezed magnon eigenstates in the qubit ground and excited states as $\ket{n}_g$ and $\ket{n}_e$, respectively. In the previous section, Eq. (\ref{squeeze}) implies that the squeezed magnon states are `squeezed' with respect to the Fock magnon states. %It was shown in Ref.~\cite{romling2023resolving} that the corresponding squeezed-magnon vacuum states, $\ket{0}_g$ and $\ket{0}_e$, are related by an effective squeezing transformation characterized by the squeezing parameter $r_{\mathrm{eff}} = r_g - r_e$, where $r_g$ and $r_e$ are the squeezing parameters associated with $\ket{n}_g$ and $\ket{n}_e$ (see Fig.~\ref{fig:system}(b) and {\color{blue}SM}). 
%Here, we demonstrate that this relation extends beyond the vacuum and holds for all eigenstates of the system. 
Using that relation and some basic operator algebra, it can be shown that the squeezed magnons in the qubit's ground state are also `squeezed' with respect to the ones in the qubit's excited state in a similar fashion (see SM [\ref{SM}] for details):
\begin{equation}\label{eq:eff_squeez}
    \ket{n}_g = \hat{S}(\xi_{\mathrm{eff}})\ket{n}_e
    = \sum_m {C}_{m,n}(\xi_{\mathrm{eff}})\ket{m}_e,
\end{equation}
where $\hat{S}(\xi_{\text{eff}})$ is the effective squeezing operator with squeezing factor of $\xi_{\text{eff}} = r_\text{eff}~ e^{i\theta}$and $r_\text{eff}=r_g-r_e$ is the difference between the squeezing parameters in the ground and excited states of the qubit. From Eqs. (\ref{squeeze}) and (\ref{eq:eff_squeez}), we infer that the information about the magnonic superpositions contained in the squeezed states gets transferred between the qubit states through the dispersive coupling. The information about the superpositions can be extracted by manipulating the state of the qubit. We discuss it in details in the following section. %Equation (\ref{eq:eff_squeez}) can be explicity written in the matrix form as

\section{Detecting squeezed magnons via qubit spectroscopy}\label{Spectroscopy}

%The qubit contains the information about the magnon-number statistics  because of the dispersive interaction. This information is accessed by performing a 
%The qubit spectroscopy involves driving the qubit with a microwave and measuring a response which is proportional to the excitation of the qubit. The qubit frequencies can be simply probed by performing a spectroscopy.
The magnonic superpositions can be resolved via a qubit spectroscopy.
In the spectroscopy, the qubit is driven by a weak ac signal and its response is recorded as a function of the drive frequency. The response peaks when the drive frequency is equal to the qubit frequency, which is the difference between its ground  and excited state energies.
%Due to its dispersive coupling with the magnon mode, the qubit also has squeezed magnon numbered levels in its ground  and excited states. This renders the qubit frequency multivalued. 
Since there can be transitions between the different squeezed magnon levels in the ground state of the qubit to those in its excited state, the qubit frequency becomes multivalued. %The dispersive coupling with the magnon mode renders the qubit frequency multivalued due to the presence of squeezed magnon levels.} 
The qubit frequency for a pair of levels $(\ket{n}_g,\ket{m}_e)$ is evaluated as (see SM [\ref{SM}] for details)
\begin{equation}\label{omega_nm}
\begin{aligned}
    \omega_{mn} = m\,\omega_{\alpha}^{e}
    - n\,\omega_{\alpha}^{g}
    + \omega_q
    - \chi
    + \frac{\omega_{\alpha}^{e} - \omega_{\alpha}^{g}}{2}.
\end{aligned}
\end{equation}
 
%However, in the linear response regime, only those transitions are allowed for which $\ket{m}_e$ has a finite overlap with $\ket{n}_g$, i.e. $C_{m,n}(\chi_\text{eff})\neq0$ [see Eq. (\ref{eq:eff_squeez})]. 

\begin{figure}[!ht]
    \centering
    \includegraphics[width=\linewidth]{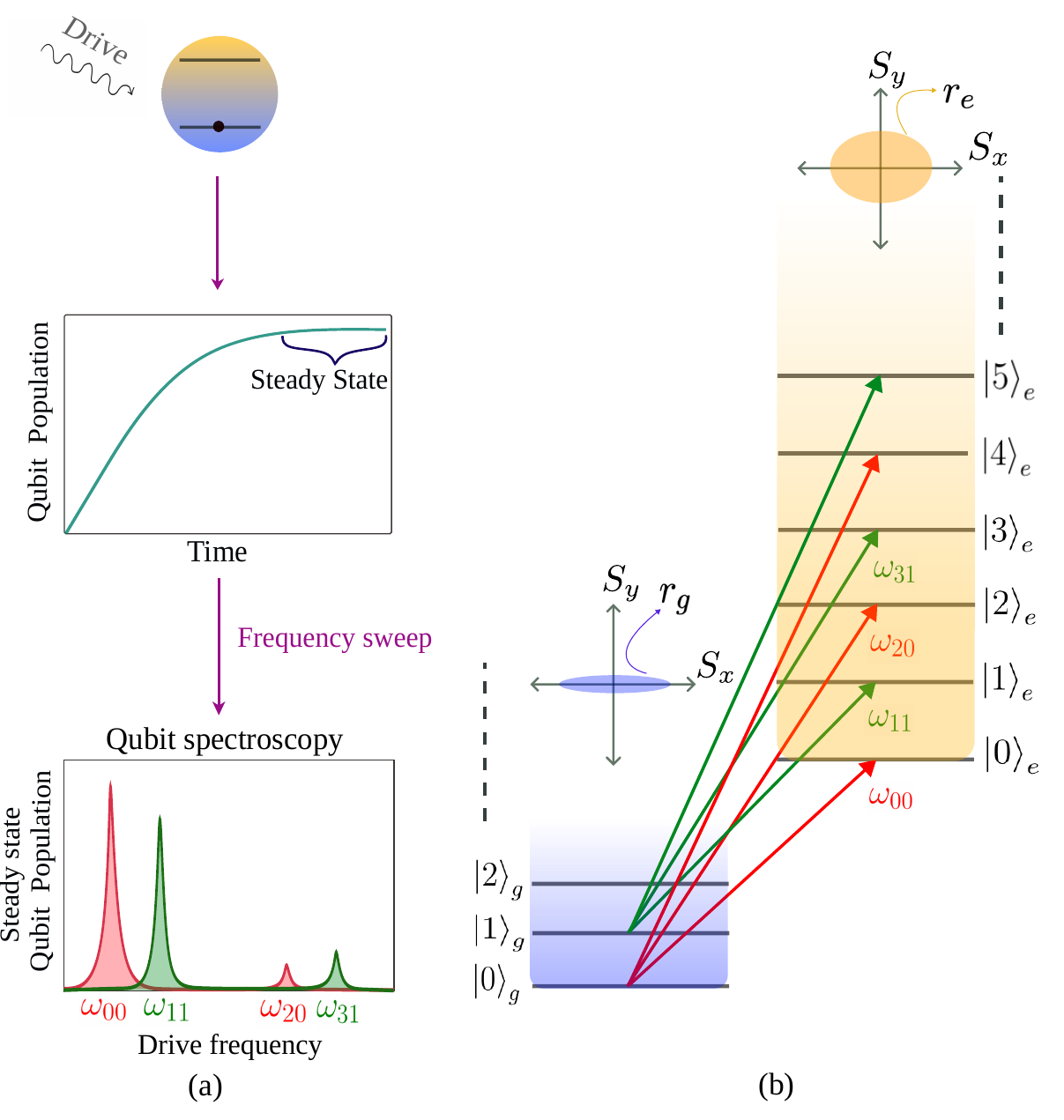}
    \caption{(a) Simulation protocol of the qubit spectroscopy. The qubit is driven by an external microwave field with amplitude \(\Omega_{d}\) and frequency \(\omega_{d}\) and is allowed to evolve until the steady state is reached. Sweeping \(\omega_{d}\) yields peaks at resonant frequencies corresponding to transitions between squeezed-magnon levels in the ground and excited states of the qubit. (b) Illustration of the system eigenstates and the transitions between them. Transitions from \(\ket{0}_{g}\) to \(\ket{0}_{e}, \ket{2}_{e}, \ldots\) (red arrows) produce the corresponding red peaks at \(\omega_{d}=\omega_{00}, \omega_{20}, \ldots\); analogous transitions from \(\ket{1}_{g}\) to \(\ket{1}_{e}, \ket{3}_{e}, \ldots\) (green arrows) yield green peaks at \(\omega_{d}=\omega_{11}, \omega_{31}, \ldots\). The set of peaks emerging due to transition from a given squeezed magnon excitation $\ket{n}_g$ state is the direct signature of the set of magnon number states whose quantum superposition constitutes $\ket{n}_g$.}
\label{fig:Qs_protocol}
\end{figure}

%----------------------------------------------------------------------------------- 
The information about the different magnon number states can be extracted by harnessing the multivalued nature of the qubit frequency. %To resolve the magnonic superpositions in $|n\rangle_g$, the system is initialized in this state. 
Let the magnon mode be in the squeezed magnon state $\ket{n}_g$ which is to be sensed.  Then, upon coherently driving the qubit with a frequency $\omega_d$, a resonant transition $\ket{n}_g\to\ket{m}_e$ occurs at $\omega_d=\omega_{mn}$ if $\ket{m}_e$ has a finite overlap ($C_{m,n}$) with $\ket{n}_e$. This gives rise to a distinct set of peaks corresponding to all the $\ket{m}_e$ states contained in the superposition [Eq.~\eqref{eq:eff_squeez}]. For a small excitation (linear response), the height of the peak at $\omega_{mn}$ is proportional to the squared overlap coefficients $|C_{m,n}(\xi_{\mathrm{eff}})|^2$  ~\cite{Roemling2023}. Consequently, the squeezed magnon state can be reconstructed by extracting the coefficients from the spectroscopic peaks and fitting them with the exact analytical expressions derived in Sec. (\ref{superpositions}). Hence, each squeezed magnon state has a distinct spectroscopic fingerprint endowed by the quantum superpositions, thereby enabling its readout. % qubit spectroscopy signal directly encodes information about the underlying magnonic state, which can be quantitatively extracted using the exact analytical expressions derived for the squeezed eigenstates. 
Figure \ref{fig:Qs_protocol} schematically depicts the qubit spectroscopy approach for resolving squeezed magnon states. The red and green colored peaks represent the magnonic statistics of the squeezed magnon vacuum and squeezed magnon single excitations, respectively.

It is to be noted that the squeezed magnon number is not conserved during the qubit excitation. This makes our sensing protocol different from the quantum nondemolition measurements performed for resolving the number levels of a given bosonic mode\cite{Schuster2007,Arrangoiz-Arriola2019,Quirion2017}.

\subsection{Qubit spectroscopy via master equation}\label{simulation}
%We use \texttt{QuTiP} package \cite{JOHANSSON20131234} to simulate the qubit spectroscopy. %The goal is to model the dynamics of a qubit which is dispersively coupled to an anisotropic ferromagnet or more precisely a magnon mode.

We simulate the spectroscopy by considering the model to be an open quantum system weakly coupled to a thermal bath. %Since we are mainly interested in assessing the efficiency of the qubit as the sensor,
The coherent drive on the qubit is modeled as $ \hat{V}_d=\Omega_{d} \cos(\omega_{d} t) (\hat{\sigma}_{+} + \hat{\sigma}_{-})$ where $\Omega_d$ signifies the strength of the drive and $\omega_d$ is the drive frequency. %Here, $\hat{\sigma}_{+}$ and $\hat{\sigma}_{-}$ are the raising and lowering operators for the qubit, respectively.
At zero temperature, the dynamics of the system is studied using the Lindblad master equation $(\hbar =1)$ \cite{Breuer}
\begin{equation}\label{Lindblad}
      \frac{d\rho}{dt}=-i[\hat{H}_\text{sys}+\hat{V}_d,\rho] + \kappa_q \left(\sigma_- \rho \sigma_+ -\frac{1}{2}\left\{\sigma_+\sigma_- , \rho \right\}\right).
 \end{equation}
where $\rho$ is the density matrix of the magnon-qubit system and $\kappa_q$ is the decay rate of the qubit. The system is initialized in the squeezed magnon state which is to be probed. The response of the qubit is obtained by computing its excitation $\langle\sigma_+\sigma_- \rangle=\text{Tr}(\rho \sigma_+\sigma_-)$ at steady state and plotting it as a function of $\omega_d$. The peaks in $\langle\sigma_+\sigma_- \rangle$ resolve the magnonic superpositions present in the squeezed magnon state. %Since we mainly want to demonstrate the probing by the qubit, we have not included magnon decay explicitly in the master equation as it only plays a role in generation of the magnon state to be probed and does not impact the qubit response. %The state in which the system is initialized implicitly takes into account this decay. 
%In a later section, we discuss a simulation in which we explicitly include the magnon decay to generate an excited squeezed magnon state from vacuum and perform its readout via the qubit. %In the {\color{blue}SM}, we have analyzed the case of finite temperature as well.
 
 In the following subsection and in Sec. \ref{detection}, we only present the results at zero temperature to demonstrate that our sensing protocol is fundamentally valid. The effect of finite temperature in the sensing is briefly discussed in Sec. \ref{limitations}.

\begin{figure}[t]
\centering
\includegraphics[width=0.9\columnwidth]{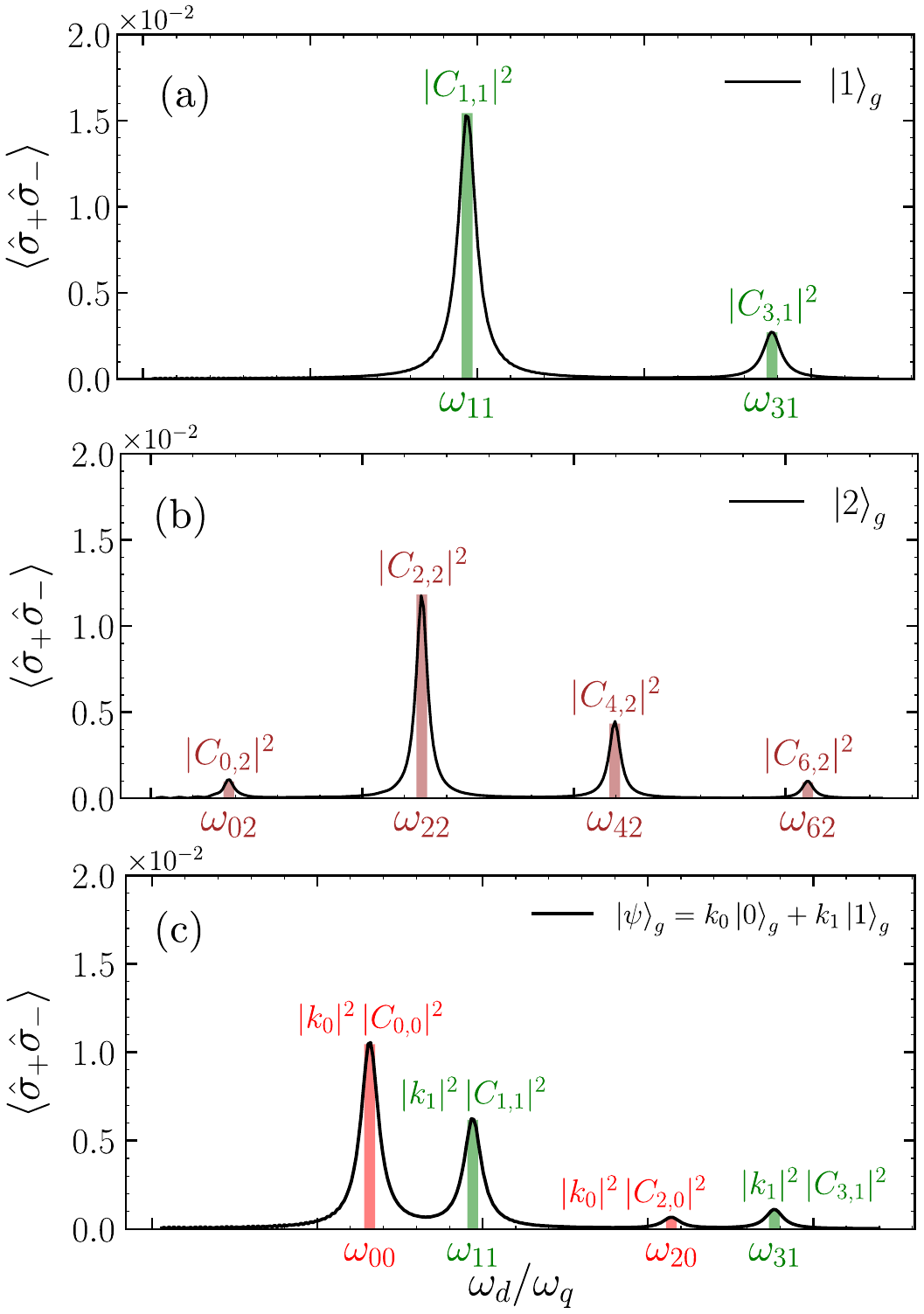}
\caption{
Steady state qubit occupation as a function of frequency of the drive for detecting different squeezed-magnon excited states. The black curve refers to the qubit occupation while the colored bars represent the analytical values of magnon number statistics calculated using Eqs. (\ref{coeff0}), (\ref{coeff1}) and (\ref{coeff2}).
(a) Detection of squeezed-magnon single-excitation $\ket{1}_{g}$.   
Peaks appear at frequencies $\omega_{11}$ and $\omega_{31}$ with heights
proportional to the analytical probabilities
$|{C}_{1,1}(r_{\mathrm{eff}})|^{2}$ and
$|{C}_{3,1}(r_{\mathrm{eff}})|^{2}$, respectively.
(b) Detection of squeezed-magnon double-excitation $\ket{2}_{g}$.  
Peaks appear at frequencies $\omega_{02},~\omega_{22},~\omega_{4,2},~\omega_{6,2}$ with heights
proportional to the analytically calculated probabilities
$|{C}_{0,2}(r_{\mathrm{eff}})|^{2},~ $ 
$|{C}_{2,2}(r_{\mathrm{eff}})|^{2},~ $
$|{C}_{4,2}(r_{\mathrm{eff}})|^{2},$ and
$|{C}_{6,2}(r_{\mathrm{eff}})|^{2}$, respectively.
(c) Detection of an arbitrary superposition state of squeezed magnon vacuum and its single excitation, $\ket{\psi}_g=k_0 |0\rangle_g+k_1 |0\rangle_g$:  
The peak heights $\propto |k_0|^2|C_{m,0}|^2$ and  $|k_1|^2|C_{m,1}|^2$  where $m=0,1,\dots$.
The simulation parameters are $r=0.3$,
$\omega_{\alpha}/\omega_{q}=0.5$,
$\chi/\omega_{q}=0.2$,
$\kappa_{q}/\omega_{q}=0.1$, and
$\Omega_{d}/\omega_{q}=0.014$,
chosen to match YIG parameters~\cite{Roemling2023}. Here $k_{0}=\sqrt{3/5}$ and $k_{1}=\sqrt{2/5}$,  but any other combination works too.
%For improved peak resolvability in (c), a qubit decay rate $\kappa_{q}=0.075\,\omega_{q}$ (instead of $0.1\,\omega_{q}$) is used to narrow the linewidth.
}
\label{fig:qs_1g_2g}
\end{figure}

\subsection{Simulation results}\label{Simulation-results }

 The spectroscopic simulation, discussed above, has been shown to resolve the magnonic superpositions in a squeezed magnon vacuum $\ket{0}_g$\cite{Roemling2023}.  This leaves an open question whether this protocol succeeds in resolving the intrinsic superpositions in an arbitrary squeezed magnon excitation $\ket{n}_g$.
Here, we show the simulation results for resolving the superpositions of the excited states $\ket{1}_{g}$ and $\ket{2}_{g}$ in Fig. \ref{fig:qs_1g_2g}. The black curve shows the numerically computed qubit excitation, while the red colored bars indicate the analytically calculated probabilities $|{C}_{m,n}(r_{\text{eff}})|^{2}$ (upto a proportionality constant independent of $m $ or $n$) which refer to the squared overlaps between the states $\ket{n_g}$ and $\ket{m_e}$ [see Eq. (\ref{eq:eff_squeez})]. For instance, when the system is prepared in the state $\ket{1}_{g}$, peaks are observed at $\omega_{d}=\omega_{11},\omega_{31},\ldots$, the heights of which are proportional to $|{C}_{1,1}(r_{\text{eff}})|^{2},|{C}_{3,1}(r_{\text{eff}})|^{2},\ldots$ respectively [Fig. \ref{fig:qs_1g_2g}(a)]. The peaks correspond to the resonant excitations $\ket{1}_g\to|1\rangle_e, \ket{1}_g\to|3\rangle_e,\ldots$. This is consistent with the fact that a single squeezed magnon excitation is superposition of odd magnon numbered Fock states. The value of the ratio $|{C}_{1,1}(r_{\text{eff}})/{C}_{3,1}(r_{\text{eff}})|^{2}$ obtained from the analytical result is in good agreement with the ratio of the corresponding peak heights obtained in the simulation within a $3\%$ relative error. Figure ~\ref{fig:qs_1g_2g}(b) shows the simulation results to resolve the superpositions in $\ket{2}_{g}$. Here, the peaks are obtained at $\omega_{d}=\omega_{02},\omega_{22}, \omega_{42},\ldots$ and their heights are proportional to $|{C}_{0,2}(r_{\text{eff}})|^{2},|{C}_{2,2}(r_{\text{eff}})|^{2},|{C}_{4,2}(r_{\text{eff}})|^{2}\ldots$ respectively.   %Figure ~\ref{fig:qs_1g_2g}(b) shows the qubit spectroscopy for the state $\ket{2}_{g}$ which consists of following even superpositions of states $\ket{2n}_{e}$ Eq.\ref{eq.2g} (derived in {\color{blue}SM}),
%\begin{equation}\ket{2}_{g} = {C}_{0,2} \ket{0}_{e} + {C}_{2,2} \ket{2}_{e} + {C}_{4,2} \ket{4}_{e} +  {C}_{6,2} \ket{6}_{e} + \ldots\label{eq.2g}\end{equation}The solid black simulation curve matches the analytical peak heights calculated using the corresponding $|{C}_{2m,2}(r_{\text{eff}})|^{2}$ coefficients. %It is worth noting that the state $\ket{2}_{g}$ contains contribution from state $\ket{0}_{e}$ as well along with the other even contribution of the states $\ket{2m}_{e}$. 
%These results show that qubit spectroscopy protocol successfully resolves the squeezed-magnon single and double excitations. 
In the similar fashion, the superpositions contained in any higher excitation $|n\rangle_g$ can also be resolved. This shows that the qubit spectroscopy can detect any squeezed magnon excitation of an anisotropic magnet.%Thus we conclude that any squeezed-magnon eigenstates can be resolved with a dispersively coupled spin qubit to an anisotropic ferromagnet.

Now we investigate whether the spectroscopic protocol can sense only an individual squeezed magnon number state (eigen-excitation) or a superposition state of different squeezed magnons.
An equal amplitude quantum superposition of a single magnon and vacuum  $\left(\ket{0}+\ket{1}\right)/\sqrt{2}$ has been observed recently using a superconducting qubit \cite{Wang2023}. It would be interesting to see whether a similar superposition of squeezed magnon vacuum and a single squeezed magnon excitation can be detected via the qubit spectroscopy. %Here, we show that it is also possible to detect 
We consider an arbitrary superposition of eigenstates $\ket{0}_{g}$ and $\ket{1}_{g}$, which can be written as
\begin{equation}
    \begin{aligned}
         \ket{\psi}_{g} &= k_{0} \ket{0}_{g} + k_{1} \ket{1}_{g} \\
         &= k_{0} \sum_{m=0}^{\infty} {C}_{2m,0}(r_{\text{eff}})\ket{2m}_{e} + k_{1} \sum_{m=0}^{\infty} {C}_{2m+1,1}(r_{\text{eff}})\ket{2m+1}_{e},
    \label{psi_g}   
    \end{aligned}
\end{equation}
where $k_{0}$ and $k_{1}$ are the respective probability amplitudes. %Figure \ref{fig:1g2gsuperposition} (a) shows the level diagram with the relevant transition along with the spectroscopy results. 
Such a state should give two sets of peaks in the spectroscopy -- one at $\omega_d=\{\omega_{00},\omega_{20}, \omega_{40},\ldots\}$ and the other at $\omega_d=\{\omega_{11},\omega_{31}, \omega_{51},\ldots\}$  --  corresponding to the  transitions from the states $\ket{0}_g$ and $\ket{1}_g$ respectively. Since the frequencies of both sets are distinct, the peaks of each set are well resolved. Due to the different weights of the two  squeezed states, the proportionality relation of the peaks with the coefficients $|C_{mn}|^2$ gets renormalized by the corresponding weights. The peak heights for the set corresponding to $\ket{0}_g$ is now proportional to $ |k_0|^2|{C}_{2m,0}(r_{\text{eff}})|^2$ while that for $\ket{1}_g$ is proportional to $|k_1|^2|{C}_{2m+1,1}(r_{\text{eff}})|^2$ where $m=0,1,2\dots$. %The spectroscopic results are shown for $k_{0} = \sqrt{\frac{3}{5}}$ and $k_{1} = \sqrt{\frac{2}{5}}$ specifically. 
As seen from Fig.~\ref{fig:qs_1g_2g}(c), the simulation results shown by solid black curve match with the analytical peak heights shown by different colored-bars corresponding to specific transitions from the two different squeezed magnon states. The agreement between the simulation and analytical results also remains good for the values of $k_0$ and $k_1$ other than the one used in the figure. This shows that the spectroscopy can resolve an arbitrary quantum superposition of $\ket{0}_g$ and $\ket{1}_g$.

By performing similar simulations, we find that the spectroscopy also resolves superpositions of $\ket{0}_g$ and $\ket{2}_g$, $\ket{1}_g$ and $\ket{3}_g$ and even the superpositions of more than two squeezed magnon states (see SM [\ref{SM}] for details). However, the irregularity in the oscillations about the mean steady state value increases for these superpositions. This introduces more numerical error in the spectrum, owing to which the proportionality of the magnonic distribution with some of the peaks is slightly affected, although the peaks still occur at the expected frequencies $\omega_{mn}$. %This implies that the master equation (\ref{Lindblad}) is  not sufficiently adequate in simulating the sensing of a generic superposition of the squeezed magnon states. 

\begin{figure}[t]
    \centering
\hspace{-0.2cm}\includegraphics[trim={0cm 0cm 0cm 0cm},clip,width=8.5cm]{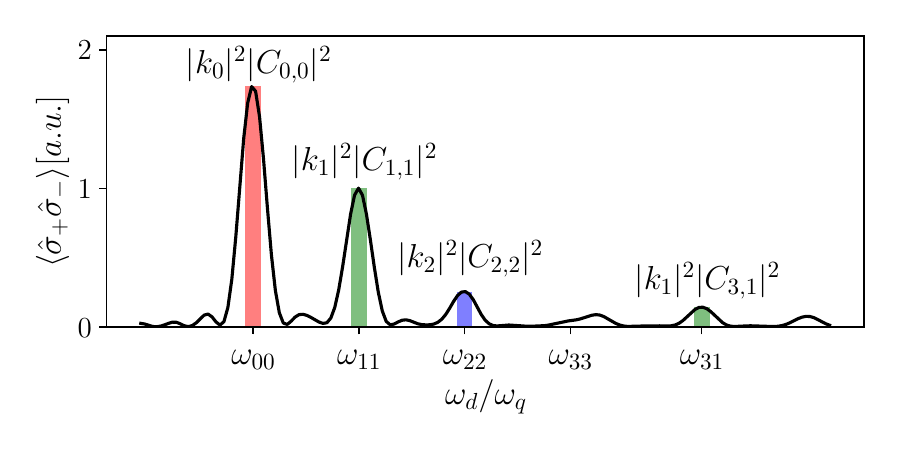}
    \caption{Qubit excitation as a function of the qubit drive frequency for the sensing of a coherent superposition of squeezed magnons generated by external driving of the magnon mode. The spectroscopic peak heights are consistent with the expected  theoretical results and still resolve the magnonic distribution. This validates the effectiveness of the sensing protocol for a deterministically prepared squeezed magnon state. In addition, it also demonstrates that the qubit spectroscopy can detect a superposition of multiple squeezed magnon number states $\ket{0}_g, \ket{1}_g$, $\ket{2}_g,\dots$ and driving the magnonic mode does not compromise the sensing by the qubit. The parameters used in the simulation are: $N_\alpha=35$, $A=0.2~\omega_q$, $\chi=0.05~\omega_q$, $B=\omega_q/16$, $\Omega_d=\omega_q/40000$, $\Omega_m=0.002 ~A$, $\kappa_m=0.0005~\omega_q$, $\kappa_q=0.00075~\omega_q$, where $N_\alpha$ is the dimension of the truncated magnon Hilbert space.}
\label{coherent_drive}
\end{figure}

\section{Generation and detection of a coherent superposition of squeezed magnons}\label{detection}

In the previous section, we considered a simplified master equation (\ref{Lindblad}) where the decay occurs only in the qubit and not in the magnon mode. This approach was taken with the expectation that the magnon decay mainly plays a role in the stabilization of the steady magnon state to be probed and does not impact the sensing by the qubit significantly. In this section, we validate this expectation and simplified analysis via a rigorous accounting of the magnon dissipation. We discuss the results of the spectroscopic simulation by including both magnon and qubit decays explicitly in the master equation.

In a typical experimental setup pertaining to such sensing protocols, nonequilibrium magnonic superpositions are achieved using an external drive, followed by the qubit spectroscopy~\cite{Quirion2017,Xu2023_magnon,Schuster2007}carried out by recording the reflection or transmission of a short pulse. In contrast, the numerical simulation protocol employed above and in previous works~\cite{Roemling2023,Dey2025} emulates the qubit spectroscopy by evaluating the qubit steady state occupation as a function of the qubit drive frequency. This protocol, while being very different from the pulse-based experiments, was found to be effective when only the qubit is our dynamical system and the magnon mode is treated as rigid. This is no longer the case when dissipation and dynamics of both the qubit and the magnonic mode are considered, which we now do. Among the key reasons for the `steady state protocol' not working, we note that the combined quantum system has joint decay channels which necessary affect both the qubit and magnon at the same time~\cite{Tarek}. Furthermore, the driven qubit exerts sufficiently large back-action on the magnonic mode such that it exerts an undesired perturbation on the latter. Thus, we now emulate the qubit spectroscopy in two steps: (i) the magnon drive is tuned on and steady state is achieved, and then (ii) the qubit drive is turned on and qubit excitation is recorded shortly after and well before the qubit steady state can be achieved. This is equivalent to probing the qubit with a short pulse that executes the spectroscopy in a noninvasive manner much like the typical experimental procedures. Please see the SM [\ref{SM}] for further details on the spectroscopy.

%Magnon decay plays a crucial role in stabilizing these states against the  external drive. Unlike the previously {\color{magenta} evaluated solutions to the master equation (Fig. \ref{fig:qs_1g_2g})} where we have implicitly taken into account the magnon decay by initializing the system in a given squeezed magnon state by default, here we incorporate  the magnon decay explicitly in the master equation and generate a nonequilibrium state of squeezed magnons. {\color{blue}We excite the magnon mode from ground state of the system in order to create a desired squeezed magnon state and perform the qubit spectroscopy on it. Such an analysis calls for a new set of collapse operators in the Lindblad equation which accounts for all the decay channels in the system. The new decay channels and the modified master equation have been briefly discussed in the SM [\ref{SM}]. Their detailed derivation does not belong here as it forms the premise of an ongoing work by one of the coauthors\cite{Tarek}. Owing to multiple decay channels now, a steady state response of the qubit does not sense the superpositions correctly. So the response is now calculated in a transient state of the qubit at a time scale which is much shorter than its relaxation time $1/\kappa_q$.}

Now, we present the simulation data for a coherent state of squeezed magnons $|\psi\rangle=\sum_{n}k_n(\beta)|n\rangle_g$. Here, $\beta$ is the coherent amplitude determined by strength of the driving and magnon decay, and $k_n(\beta)=e^{-\frac{|\beta|^2}{2}}\frac{\beta^n}{\sqrt{n!}}$ is the probability amplitude of the squeezed magnon number states.  The state is prepared as follows. Since the system is initially at equilibrium with the bath at zero temperature,  it is in the ground state $\ket{0}_g$. So, we start with the state $\ket{0}_g$ in the master equation. The magnon mode is then resonantly driven with an ac field at a frequency $\omega =\omega_\alpha^g$. This should excite the squeezed magnon mode in the ground state of the qubit into its coherent state $\ket{\psi}$.  In the master equation, the magnon drive is modeled as $\Omega_m \cos \omega_\alpha^g t~ ( \hat{\alpha}_g + \hat{\alpha}_g^\dagger)$ where $\Omega
_m$ denotes the driving amplitude and the magnon decay rate is given by $\kappa_m$. The driving amplitude is kept low which populates only the lowest lying number states ($\bar{n}\approx0.64$). By evolving the system in time, we obtain a coherent state distribution of the squeezed magnons at the steady state.

The driven steady magnon state now acts as the quantum state to be probed. With the magnon drive still on, the qubit spectroscopy tone $\Omega_{d} \cos(\omega_{d} t) (\hat{\sigma}_{+} + \hat{\sigma}_{-})$ is also switched on. The qubit excitation is recorded at a short time $\Delta t\approx 0.11/\kappa_q$ elapsed from the onset of the drive. This technique is designed to emulate a reflection based spectrocopy of the qubit\cite{Quirion2017}.
 The plot of the qubit excitation as a function of driving frequency for the nonequilibrium squeezed magnon state is shown in Fig. \ref{coherent_drive}. Analogous to the case of the superposition state of vacuum and a single squeezed magnon excitation discussed in previous section, the coherent state also displays multiple sets of peaks in the qubit spectroscopy at $\omega_d=\{\omega_{mn}\}$ with heights $\{|k_n|^2 |C_{m,n}|^2\}$ for $m,n=0,1,2\dots$, due to the occupation of multiple squeezed magnon levels $(n)$. The heights of the resonant peaks (black curve) are in agreement with the expected `scaled' probability distribution (colored bars). The agreement has been shown only for the peaks which correspond to the dominant channels of excitation. The smaller peaks appearing between the dominant ones correspond to the excitations $\ket{n}_g\to\ket{m}_e$ with significantly low overlaps $\sim|C_{mn}(r_\text{eff})|^2$ or low occupation probability $|k_n|^2$ of the $\ket{n}_g$ levels. Due to proximity of their frequencies, some of these  peaks merge to form a single peak and the corresponding magnon number states cannot be resolved. However, the lower peaks do not affect the resolution of the dominant ones. Hence, the sensing of the magnonic statistics still works adequately for a nonequilibrium superposition of the squeezed magnon states.

\section{Spectral crowding and finite temperature effects}\label{limitations}

\begin{figure}[h]
    \centering
    \includegraphics[width=0.47\textwidth]{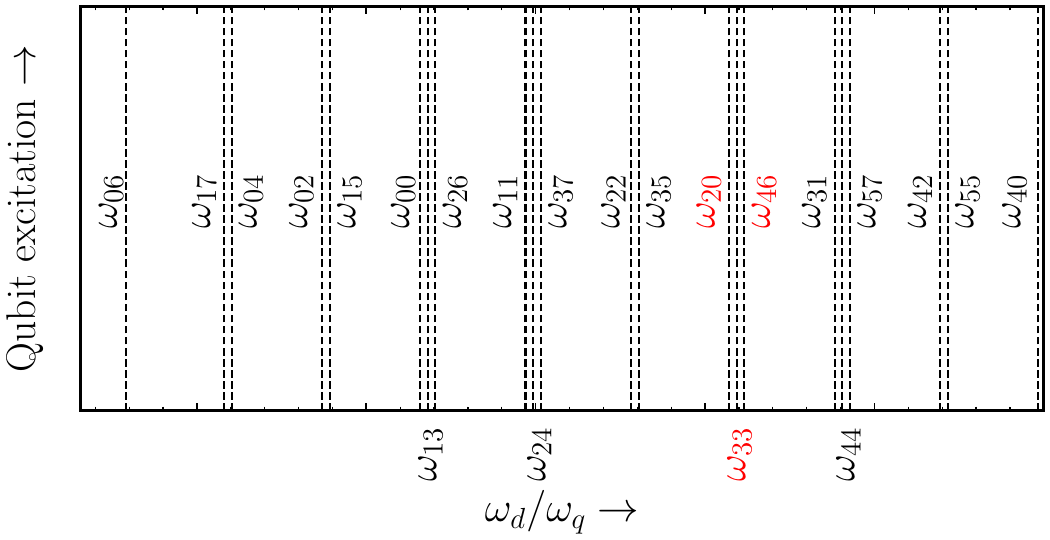}
    \caption{Resonant transition frequencies $\omega_{mn}$ (vertical dashed lines) across the full drive-frequency sweep $\omega_{d}/\omega_{q}$, where $\omega_{mn}$ denotes the transition $|n\rangle_{g}\!\to\!|m\rangle_{e}$. The clustering of distinct $\omega_{mn}$ values at higher manifolds leads to spectral crowding and overlapping resonances.  Transitions highlighted in red mark near-degenerate lines that hinder resolving superposed squeezed magnon states whose components involve these frequencies. This issue arises while detecting the coherently driven squeezed magnon state [Fig. \ref{coherent_drive}] as the peaks around $\omega_{33}$ could not be resolved.} 
\label{fig:transition_frequencies}
\end{figure}

Despite the fact that the sensing remains valid for the generated coherent state, it has some shortcomings. Some of the multivalued qubit frequencies lie very close to each other, a phenomenon known as spectral crowding [Fig. \ref{fig:transition_frequencies}]. It degrades the resolution of the peaks corresponding to the transitions at those frequencies, say $\omega_{20},~\omega_{33},~\omega_{46}$ for instance (marked red in the figure).  The coalescence of the neighboring peaks not only affects the resolution but  also compromises the proportionality of the peaks with the actual probability distribution. This issue does not jeopardize our sensing protocol if the closely-spaced peaks are low enough and occur between the peaks corresponding to the dominant excitations, as shown in Fig. \ref{coherent_drive}. However, they can be detrimental to the sensing if the magnon drive is strong enough to significantly populate the levels whose excitations correspond to the closely lying peaks. So, the spectroscopy gives better results for low intensity magnon drives. Furthermore, the frequency linewidth of the qubit, proportional to its decay rate $\kappa_q$, should be much smaller than the separation between those peaks in order to resolve the number states accurately. 

\begin{figure}[t]
    \centering
\includegraphics[width=8.5cm]{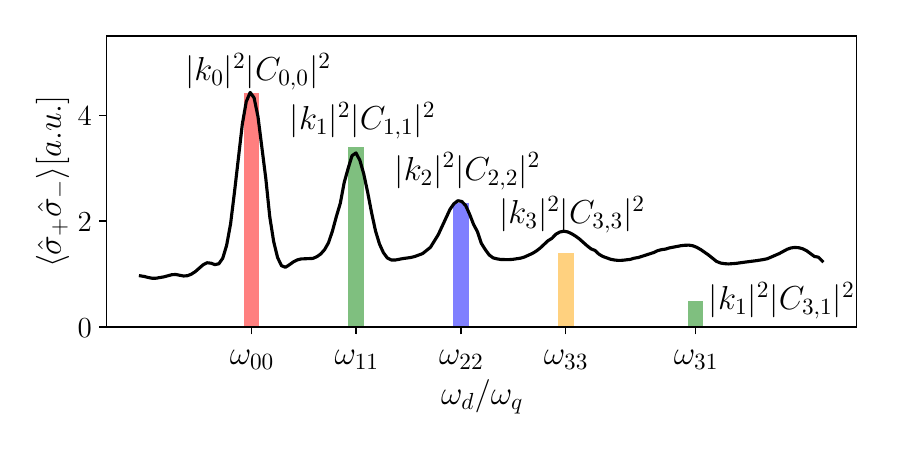}
    \caption{Qubit excitation as a function of the qubit drive frequency for sensing of a coherent superposition of squeezed magnons at $k_BT=0.5 \omega_q$. The spectroscopic peaks do not read the higher squeezed magnon number states accurately as seen from the mismatch of peak heights and the analytical probabilities (colored bars) at $\omega_{33}$ and $\omega_
    {31}$. This shows that our spectroscopic protocol does not work well at high temperatures. The simulation parameters are the same as in Fig. \ref{coherent_drive}.}
\label{coherent_drive_T}
\end{figure}

An additional challenge arises from finite-temperature effects in the system.   %Considering only the qubit to be coupled with the thermal bath, the sensing remains effective for detecting squeezed-magnon states at low temperatures, specifically upto \(k_BT \approx 0.2\,\omega_q\). 
 We follow a similar procedure as in Sec. \ref{detection} but now the generated state of the magnons contains thermal contributions in addition to the coherent state statistics. At higher temperatures, thermal occupation of the squeezed magnon levels in the excited state of the qubit also becomes significant. To see the thermal effects, we use $T\neq0$ which gives $\bar{n}(\omega)\neq0$ for different transition channels in the modified master equation (see SM [\ref{SM}]).
The simulation outcome for a moderately high temperature bath $k_BT=0.5\omega_q$ is shown in Fig. \ref{coherent_drive_T}. We observe a finite offset of the spectroscopic curve in the figure, which corresponds to the mean thermal excitation in the qubit. The finite thermal occupation of the excited states of the qubit can render the magnitude of its excitation from the drive disproportional to the population in its ground state. Although the proportionality is not adversely affected for the dominant peaks, it is compromised for the peaks which correspond to the weaker transitions (e.g. at $\omega_{33}$ and $\omega_{31}$). The activation of more channels of transitions due to their thermal occupation also interferes with the spectroscopy because of spectral crowding. Thus, the qubit spectroscopy at finite temperatures does not reach the same quality as for $T=0$  to measure the expected probability distribution of the magnonic superpositions. A bosonic probe might be a better candidate for such a sensing at finite temperatures\cite{Dey2025}. 

\section{Conclusion and outlook}\label{conclusion}

To conclude, we show how a qubit may serve as a spectroscopic probe for sensing the magnonic superpositions that form quantum states of the   composite excitations --  squeezed magnons -- of an anisotropic ferromagnet. To validate our proposal, we perform numerical simulations to demonstrate the detection of squeezed magnon single excitation and a squeezed magnon double excitation. The probing can be generalized to any $n$ squeezed magnon excitations. We also show that the spectroscopic protocol can resolve a quantum superposition state of the squeezed magnon vacuum and a single squeezed magnon excitation. To make our analysis more concrete, we perform a modified spectroscopic protocol to demonstrate the generation of a coherent state of squeezed-magnons through a magnon drive, followed by its sensing using the qubit. We have explicitly derived the analytical expressions for some of the squeezed-magnon eigenstates, enabling a direct comparison between theory and numerical simulations, which we find to be in close agreement. We have further discussed the challenges associated with the sensing protocol, in particular those arising from spectral crowding and finite temperature effects.

Our work marks an important theoretical development in the formulation of sensing protocols for investigating a general squeezed magnon state. The protocol can also be used to demonstrate that the eigen-excitations of an anisotropic ferromagnet are squeezed magnons. The fact that these states can be detected by simply observing the qubit resonances provides an attractive alternative to other protocols which detect squeezing through measurement of the magnetization fluctuations \cite{Shimizu2025,Hioki2026} or magnon-qubit Raman assisted sweeping\cite{Weng2026}.  

%Looking ahead, an important extension of the present work is the inclusion of dissipation mechanisms beyond qubit relaxation. In particular, incorporating magnon decay due to coupling of the magnon mode to a thermal bath will be essential for a more realistic description of experimental platforms and for assessing the robustness of the spectroscopy protocol under dissipative conditions.}

\section*{Supplementary material}\label{SM}

The {\color{blue} Supplementary Material} contains:
(I) a derivation of the effective squeezing between the qubit states, 
(II) a derivation of the analytical expression of squeezed magnon single  and double excitations, 
(III) derivation of the expression of the multivalued qubit frequencies arising from the disperive coupling,
(IV) a discussion on the Lindblad master equation of the full system including both qubit and magnon decays to a common thermal bath, and 
(V) a discussion of the results of qubit spectroscopy for different superpositions of squeezed magnon states.

\begin{acknowledgments}
Financial support by the DFG (German Research Foundation) via Spin+X TRR 173-268565370 (project A13) is gratefully acknowledged.
\end{acknowledgments}

\section*{AUTHOR DECLARATIONS}

\subsection*{Conflict of Interest}
The authors have no conflicts to disclose.

\subsection*{Author Contributions}

{\bf Amey S.~Rodge}: Formal analysis (lead);
Investigation (lead); Writing – original draft (lead); Writing –
review and editing (supporting). {\bf Tarek Moussa}: Formal analysis (supporting);
Investigation (supporting); Writing – original draft (supporting); Writing –
review and editing (supporting); Conceptualization
(supporting). {\bf Akashdeep Kamra}: Conceptualization
(lead); Funding acquisition (lead); Investigation (supporting); Supervision (equal); Writing – original draft (supporting); Writing – review and
editing (supporting). {\bf Bashab Dey}: Conceptualization
(equal); Investigation (lead); Supervision (equal); Writing – original draft (lead); Writing – review and
editing (lead).

\section*{Data Availability Statement}
The data that support the findings of this study are available from the corresponding author upon reasonable request.

%--------------------------------------------A1---------------------------------------------------- %

%\nocite{*}
\section*{References}
\bibliography{composite}% Produces the bibliography via BibTeX.

%aipnum4-2.bst 2019-01-14 (MD) hand-edited version of apsrev4-1.bst
\begin{thebibliography}{67}%
\makeatletter
\providecommand \@ifxundefined [1]{%
 \@ifx{#1\undefined}
}%
\providecommand \@ifnum [1]{%
 \ifnum #1\expandafter \@firstoftwo
 \else \expandafter \@secondoftwo
 \fi
}%
\providecommand \@ifx [1]{%
 \ifx #1\expandafter \@firstoftwo
 \else \expandafter \@secondoftwo
 \fi
}%
\providecommand \natexlab [1]{#1}%
\providecommand \enquote  [1]{``#1''}%
\providecommand \bibnamefont  [1]{#1}%
\providecommand \bibfnamefont [1]{#1}%
\providecommand \citenamefont [1]{#1}%
\providecommand \href@noop [0]{\@secondoftwo}%
\providecommand \href [0]{\begingroup \@sanitize@url \@href}%
\providecommand \@href[1]{\@@startlink{#1}\@@href}%
\providecommand \@@href[1]{\endgroup#1\@@endlink}%
\providecommand \@sanitize@url [0]{\catcode `\\12\catcode `\$12\catcode
  `\&12\catcode `\#12\catcode `\^12\catcode `\_12\catcode `\%12\relax}%
\providecommand \@@startlink[1]{}%
\providecommand \@@endlink[0]{}%
\providecommand \url  [0]{\begingroup\@sanitize@url \@url }%
\providecommand \@url [1]{\endgroup\@href {#1}{\urlprefix }}%
\providecommand \urlprefix  [0]{URL }%
\providecommand \Eprint [0]{\href }%
\providecommand \doibase [0]{https://doi.org/}%
\providecommand \selectlanguage [0]{\@gobble}%
\providecommand \bibinfo  [0]{\@secondoftwo}%
\providecommand \bibfield  [0]{\@secondoftwo}%
\providecommand \translation [1]{[#1]}%
\providecommand \BibitemOpen [0]{}%
\providecommand \bibitemStop [0]{}%
\providecommand \bibitemNoStop [0]{.\EOS\space}%
\providecommand \EOS [0]{\spacefactor3000\relax}%
\providecommand \BibitemShut  [1]{\csname bibitem#1\endcsname}%
\let\auto@bib@innerbib\@empty
%</preamble>
\bibitem [{\citenamefont {Holstein}\ and\ \citenamefont
  {Primakoff}(1940)}]{Holstein1940}%
  \BibitemOpen
  \bibfield  {author} {\bibinfo {author} {\bibfnamefont {T.}~\bibnamefont
  {Holstein}}\ and\ \bibinfo {author} {\bibfnamefont {H.}~\bibnamefont
  {Primakoff}},\ }\bibfield  {title} {\enquote {\bibinfo {title} {Field
  dependence of the intrinsic domain magnetization of a ferromagnet},}\ }\href
  {https://doi.org/10.1103/PhysRev.58.1098} {\bibfield  {journal} {\bibinfo
  {journal} {Phys. Rev.}\ }\textbf {\bibinfo {volume} {58}},\ \bibinfo {pages}
  {1098--1113} (\bibinfo {year} {1940})}\BibitemShut {NoStop}%
\bibitem [{\citenamefont {Yuan}\ \emph {et~al.}(2022)\citenamefont {Yuan},
  \citenamefont {Cao}, \citenamefont {Kamra}, \citenamefont {Duine},\ and\
  \citenamefont {Yan}}]{Yuan2022}%
  \BibitemOpen
  \bibfield  {author} {\bibinfo {author} {\bibfnamefont {H.}~\bibnamefont
  {Yuan}}, \bibinfo {author} {\bibfnamefont {Y.}~\bibnamefont {Cao}}, \bibinfo
  {author} {\bibfnamefont {A.}~\bibnamefont {Kamra}}, \bibinfo {author}
  {\bibfnamefont {R.~A.}\ \bibnamefont {Duine}},\ and\ \bibinfo {author}
  {\bibfnamefont {P.}~\bibnamefont {Yan}},\ }\bibfield  {title} {\enquote
  {\bibinfo {title} {Quantum magnonics: When magnon spintronics meets quantum
  information science},}\ }\href
  {https://doi.org/https://doi.org/10.1016/j.physrep.2022.03.002} {\bibfield
  {journal} {\bibinfo  {journal} {Physics Reports}\ }\textbf {\bibinfo {volume}
  {965}},\ \bibinfo {pages} {1--74} (\bibinfo {year} {2022})}\BibitemShut
  {NoStop}%
\bibitem [{\citenamefont {Rezende}\ and\ \citenamefont
  {Zagury}(1969)}]{Rezende1969}%
  \BibitemOpen
  \bibfield  {author} {\bibinfo {author} {\bibfnamefont {S.}~\bibnamefont
  {Rezende}}\ and\ \bibinfo {author} {\bibfnamefont {N.}~\bibnamefont
  {Zagury}},\ }\bibfield  {title} {\enquote {\bibinfo {title} {Coherent magnon
  states},}\ }\href
  {https://doi.org/https://doi.org/10.1016/0375-9601(69)90788-9} {\bibfield
  {journal} {\bibinfo  {journal} {Physics Letters A}\ }\textbf {\bibinfo
  {volume} {29}},\ \bibinfo {pages} {47--48} (\bibinfo {year}
  {1969})}\BibitemShut {NoStop}%
\bibitem [{\citenamefont {Sharma}\ \emph {et~al.}(2021)\citenamefont {Sharma},
  \citenamefont {Bittencourt}, \citenamefont {Karenowska},\ and\ \citenamefont
  {Kusminskiy}}]{Sharma2021}%
  \BibitemOpen
  \bibfield  {author} {\bibinfo {author} {\bibfnamefont {S.}~\bibnamefont
  {Sharma}}, \bibinfo {author} {\bibfnamefont {V.~A. S.~V.}\ \bibnamefont
  {Bittencourt}}, \bibinfo {author} {\bibfnamefont {A.~D.}\ \bibnamefont
  {Karenowska}},\ and\ \bibinfo {author} {\bibfnamefont {S.~V.}\ \bibnamefont
  {Kusminskiy}},\ }\bibfield  {title} {\enquote {\bibinfo {title} {Spin cat
  states in ferromagnetic insulators},}\ }\href
  {https://doi.org/10.1103/PhysRevB.103.L100403} {\bibfield  {journal}
  {\bibinfo  {journal} {Phys. Rev. B}\ }\textbf {\bibinfo {volume} {103}},\
  \bibinfo {pages} {L100403} (\bibinfo {year} {2021})}\BibitemShut {NoStop}%
\bibitem [{\citenamefont {Kamra}\ and\ \citenamefont
  {Belzig}(2016{\natexlab{a}})}]{Kamra2016}%
  \BibitemOpen
  \bibfield  {author} {\bibinfo {author} {\bibfnamefont {A.}~\bibnamefont
  {Kamra}}\ and\ \bibinfo {author} {\bibfnamefont {W.}~\bibnamefont {Belzig}},\
  }\bibfield  {title} {\enquote {\bibinfo {title} {Super-poissonian shot noise
  of squeezed-magnon mediated spin transport},}\ }\href
  {https://doi.org/10.1103/PhysRevLett.116.146601} {\bibfield  {journal}
  {\bibinfo  {journal} {Phys. Rev. Lett.}\ }\textbf {\bibinfo {volume} {116}},\
  \bibinfo {pages} {146601} (\bibinfo {year} {2016}{\natexlab{a}})}\BibitemShut
  {NoStop}%
\bibitem [{\citenamefont {Yuan}\ and\ \citenamefont
  {Duine}(2020)}]{Yuan2020_antibunch}%
  \BibitemOpen
  \bibfield  {author} {\bibinfo {author} {\bibfnamefont {H.~Y.}\ \bibnamefont
  {Yuan}}\ and\ \bibinfo {author} {\bibfnamefont {R.~A.}\ \bibnamefont
  {Duine}},\ }\bibfield  {title} {\enquote {\bibinfo {title} {Magnon
  antibunching in a nanomagnet},}\ }\href
  {https://doi.org/10.1103/PhysRevB.102.100402} {\bibfield  {journal} {\bibinfo
   {journal} {Phys. Rev. B}\ }\textbf {\bibinfo {volume} {102}},\ \bibinfo
  {pages} {100402(R)} (\bibinfo {year} {2020})}\BibitemShut {NoStop}%
\bibitem [{\citenamefont {Glauber}(1963)}]{Glauber1963}%
  \BibitemOpen
  \bibfield  {author} {\bibinfo {author} {\bibfnamefont {R.~J.}\ \bibnamefont
  {Glauber}},\ }\bibfield  {title} {\enquote {\bibinfo {title} {Coherent and
  incoherent states of the radiation field},}\ }\href
  {https://doi.org/10.1103/PhysRev.131.2766} {\bibfield  {journal} {\bibinfo
  {journal} {Phys. Rev.}\ }\textbf {\bibinfo {volume} {131}},\ \bibinfo {pages}
  {2766--2788} (\bibinfo {year} {1963})}\BibitemShut {NoStop}%
\bibitem [{\citenamefont {Gerry}\ and\ \citenamefont
  {Knight}(2005)}]{Gerry2005}%
  \BibitemOpen
  \bibfield  {author} {\bibinfo {author} {\bibfnamefont {C.~C.}\ \bibnamefont
  {Gerry}}\ and\ \bibinfo {author} {\bibfnamefont {P.~L.}\ \bibnamefont
  {Knight}},\ }\href
  {https://www.cambridge.org/core/books/introductory-quantum-optics/B9866F1F40C45936A81D03AF7617CF44}
  {\emph {\bibinfo {title} {Introductory Quantum Optics}}}\ (\bibinfo
  {publisher} {Cambridge University Press},\ \bibinfo {address} {Cambridge},\
  \bibinfo {year} {2005})\BibitemShut {NoStop}%
\bibitem [{\citenamefont {Lachance-Quirion}\ \emph {et~al.}(2017)\citenamefont
  {Lachance-Quirion}, \citenamefont {Tabuchi}, \citenamefont {Ishino},
  \citenamefont {Noguchi}, \citenamefont {Ishikawa}, \citenamefont {Yamazaki},\
  and\ \citenamefont {Nakamura}}]{Quirion2017}%
  \BibitemOpen
  \bibfield  {author} {\bibinfo {author} {\bibfnamefont {D.}~\bibnamefont
  {Lachance-Quirion}}, \bibinfo {author} {\bibfnamefont {Y.}~\bibnamefont
  {Tabuchi}}, \bibinfo {author} {\bibfnamefont {S.}~\bibnamefont {Ishino}},
  \bibinfo {author} {\bibfnamefont {A.}~\bibnamefont {Noguchi}}, \bibinfo
  {author} {\bibfnamefont {T.}~\bibnamefont {Ishikawa}}, \bibinfo {author}
  {\bibfnamefont {R.}~\bibnamefont {Yamazaki}},\ and\ \bibinfo {author}
  {\bibfnamefont {Y.}~\bibnamefont {Nakamura}},\ }\bibfield  {title} {\enquote
  {\bibinfo {title} {Resolving quanta of collective spin excitations in a
  millimeter-sized ferromagnet},}\ }\href
  {https://doi.org/10.1126/sciadv.1603150} {\bibfield  {journal} {\bibinfo
  {journal} {Science Advances}\ }\textbf {\bibinfo {volume} {3}},\ \bibinfo
  {pages} {e1603150} (\bibinfo {year} {2017})}\BibitemShut {NoStop}%
\bibitem [{\citenamefont {Lachance-Quirion}\ \emph {et~al.}(2020)\citenamefont
  {Lachance-Quirion}, \citenamefont {Wolski}, \citenamefont {Tabuchi},
  \citenamefont {Kono}, \citenamefont {Usami},\ and\ \citenamefont
  {Nakamura}}]{Quirion2020}%
  \BibitemOpen
  \bibfield  {author} {\bibinfo {author} {\bibfnamefont {D.}~\bibnamefont
  {Lachance-Quirion}}, \bibinfo {author} {\bibfnamefont {S.~P.}\ \bibnamefont
  {Wolski}}, \bibinfo {author} {\bibfnamefont {Y.}~\bibnamefont {Tabuchi}},
  \bibinfo {author} {\bibfnamefont {S.}~\bibnamefont {Kono}}, \bibinfo {author}
  {\bibfnamefont {K.}~\bibnamefont {Usami}},\ and\ \bibinfo {author}
  {\bibfnamefont {Y.}~\bibnamefont {Nakamura}},\ }\bibfield  {title} {\enquote
  {\bibinfo {title} {Entanglement-based single-shot detection of a single
  magnon with a superconducting qubit},}\ }\href
  {https://doi.org/10.1126/science.aaz9236} {\bibfield  {journal} {\bibinfo
  {journal} {Science}\ }\textbf {\bibinfo {volume} {367}},\ \bibinfo {pages}
  {425--428} (\bibinfo {year} {2020})}\BibitemShut {NoStop}%
\bibitem [{\citenamefont {Wang}\ \emph {et~al.}(2021)\citenamefont {Wang},
  \citenamefont {Xiao}, \citenamefont {Guo}, \citenamefont {Lee-Wong},
  \citenamefont {Yan}, \citenamefont {Cheng},\ and\ \citenamefont
  {Du}}]{WangH2021}%
  \BibitemOpen
  \bibfield  {author} {\bibinfo {author} {\bibfnamefont {H.}~\bibnamefont
  {Wang}}, \bibinfo {author} {\bibfnamefont {Y.}~\bibnamefont {Xiao}}, \bibinfo
  {author} {\bibfnamefont {M.}~\bibnamefont {Guo}}, \bibinfo {author}
  {\bibfnamefont {E.}~\bibnamefont {Lee-Wong}}, \bibinfo {author}
  {\bibfnamefont {G.~Q.}\ \bibnamefont {Yan}}, \bibinfo {author} {\bibfnamefont
  {R.}~\bibnamefont {Cheng}},\ and\ \bibinfo {author} {\bibfnamefont {C.~R.}\
  \bibnamefont {Du}},\ }\bibfield  {title} {\enquote {\bibinfo {title} {Spin
  pumping of an easy-plane antiferromagnet enhanced by dzyaloshinskii--moriya
  interaction},}\ }\href {https://doi.org/10.1103/PhysRevLett.127.117202}
  {\bibfield  {journal} {\bibinfo  {journal} {Phys. Rev. Lett.}\ }\textbf
  {\bibinfo {volume} {127}},\ \bibinfo {pages} {117202} (\bibinfo {year}
  {2021})}\BibitemShut {NoStop}%
\bibitem [{\citenamefont {Rani}\ \emph {et~al.}(2025)\citenamefont {Rani},
  \citenamefont {Cao}, \citenamefont {Baptista}, \citenamefont {Hoffmann},\
  and\ \citenamefont {Pfaff}}]{Rani2025}%
  \BibitemOpen
  \bibfield  {author} {\bibinfo {author} {\bibfnamefont {S.}~\bibnamefont
  {Rani}}, \bibinfo {author} {\bibfnamefont {X.}~\bibnamefont {Cao}}, \bibinfo
  {author} {\bibfnamefont {A.~E.}\ \bibnamefont {Baptista}}, \bibinfo {author}
  {\bibfnamefont {A.}~\bibnamefont {Hoffmann}},\ and\ \bibinfo {author}
  {\bibfnamefont {W.}~\bibnamefont {Pfaff}},\ }\bibfield  {title} {\enquote
  {\bibinfo {title} {High-dynamic-range quantum sensing of magnons and their
  dynamics using a superconducting qubit},}\ }\href
  {https://doi.org/10.1103/6dmm-mnxd} {\bibfield  {journal} {\bibinfo
  {journal} {Phys. Rev. Appl.}\ }\textbf {\bibinfo {volume} {23}},\ \bibinfo
  {pages} {064032} (\bibinfo {year} {2025})}\BibitemShut {NoStop}%
\bibitem [{\citenamefont {Shimizu}\ \emph {et~al.}(2025)\citenamefont
  {Shimizu}, \citenamefont {Hioki}, \citenamefont {Takeda},\ and\ \citenamefont
  {Saitoh}}]{Shimizu2025}%
  \BibitemOpen
  \bibfield  {author} {\bibinfo {author} {\bibfnamefont {H.}~\bibnamefont
  {Shimizu}}, \bibinfo {author} {\bibfnamefont {T.}~\bibnamefont {Hioki}},
  \bibinfo {author} {\bibfnamefont {S.}~\bibnamefont {Takeda}},\ and\ \bibinfo
  {author} {\bibfnamefont {E.}~\bibnamefont {Saitoh}},\ }\bibfield  {title}
  {\enquote {\bibinfo {title} {Tomography of parametric transition in
  magnets},}\ }\href {https://doi.org/10.1103/v1rk-rtrq} {\bibfield  {journal}
  {\bibinfo  {journal} {Phys. Rev. Lett.}\ }\textbf {\bibinfo {volume} {135}},\
  \bibinfo {pages} {106701} (\bibinfo {year} {2025})}\BibitemShut {NoStop}%
\bibitem [{\citenamefont {Weng}\ \emph {et~al.}(2026)\citenamefont {Weng},
  \citenamefont {Xu}, \citenamefont {Chen}, \citenamefont {Tan}, \citenamefont
  {Gu}, \citenamefont {Li}, \citenamefont {Yu}, \citenamefont {Zhu},
  \citenamefont {Hu}, \citenamefont {Nori},\ and\ \citenamefont
  {You}}]{Weng2026}%
  \BibitemOpen
  \bibfield  {author} {\bibinfo {author} {\bibfnamefont {Y.-C.}\ \bibnamefont
  {Weng}}, \bibinfo {author} {\bibfnamefont {D.}~\bibnamefont {Xu}}, \bibinfo
  {author} {\bibfnamefont {Z.}~\bibnamefont {Chen}}, \bibinfo {author}
  {\bibfnamefont {L.-Z.}\ \bibnamefont {Tan}}, \bibinfo {author} {\bibfnamefont
  {X.-K.}\ \bibnamefont {Gu}}, \bibinfo {author} {\bibfnamefont
  {J.}~\bibnamefont {Li}}, \bibinfo {author} {\bibfnamefont {H.-F.}\
  \bibnamefont {Yu}}, \bibinfo {author} {\bibfnamefont {S.-Y.}\ \bibnamefont
  {Zhu}}, \bibinfo {author} {\bibfnamefont {X.}~\bibnamefont {Hu}}, \bibinfo
  {author} {\bibfnamefont {F.}~\bibnamefont {Nori}},\ and\ \bibinfo {author}
  {\bibfnamefont {J.~Q.}\ \bibnamefont {You}},\ }\bibfield  {title} {\enquote
  {\bibinfo {title} {Magnon squeezing in the quantum regime},}\ }\href
  {https://doi.org/10.1038/s41467-026-69312-4} {\bibfield  {journal} {\bibinfo
  {journal} {Nature Communications}\ }\textbf {\bibinfo {volume} {17}},\
  \bibinfo {pages} {2679} (\bibinfo {year} {2026})}\BibitemShut {NoStop}%
\bibitem [{\citenamefont {Laucht}\ \emph {et~al.}(2021)\citenamefont {Laucht},
  \citenamefont {Hohls}, \citenamefont {Ubbelohde}, \citenamefont {Fernando
  Gonzalez-Zalba}, \citenamefont {Reilly}, \citenamefont {Stobbe},
  \citenamefont {Schröder}, \citenamefont {Scarlino}, \citenamefont {Koski},
  \citenamefont {Dzurak}, \citenamefont {Yang}, \citenamefont {Yoneda},
  \citenamefont {Kuemmeth}, \citenamefont {Bluhm}, \citenamefont {Pla},
  \citenamefont {Hill}, \citenamefont {Salfi}, \citenamefont {Oiwa},
  \citenamefont {Muhonen}, \citenamefont {Verhagen}, \citenamefont {LaHaye},
  \citenamefont {Kim}, \citenamefont {Tsen}, \citenamefont {Culcer},
  \citenamefont {Geresdi}, \citenamefont {Mol}, \citenamefont {Mohan},
  \citenamefont {Jain},\ and\ \citenamefont {Baugh}}]{Laucht2021}%
  \BibitemOpen
  \bibfield  {author} {\bibinfo {author} {\bibfnamefont {A.}~\bibnamefont
  {Laucht}}, \bibinfo {author} {\bibfnamefont {F.}~\bibnamefont {Hohls}},
  \bibinfo {author} {\bibfnamefont {N.}~\bibnamefont {Ubbelohde}}, \bibinfo
  {author} {\bibfnamefont {M.}~\bibnamefont {Fernando Gonzalez-Zalba}},
  \bibinfo {author} {\bibfnamefont {D.~J.}\ \bibnamefont {Reilly}}, \bibinfo
  {author} {\bibfnamefont {S.}~\bibnamefont {Stobbe}}, \bibinfo {author}
  {\bibfnamefont {T.}~\bibnamefont {Schröder}}, \bibinfo {author}
  {\bibfnamefont {P.}~\bibnamefont {Scarlino}}, \bibinfo {author}
  {\bibfnamefont {J.~V.}\ \bibnamefont {Koski}}, \bibinfo {author}
  {\bibfnamefont {A.}~\bibnamefont {Dzurak}}, \bibinfo {author} {\bibfnamefont
  {C.-H.}\ \bibnamefont {Yang}}, \bibinfo {author} {\bibfnamefont
  {J.}~\bibnamefont {Yoneda}}, \bibinfo {author} {\bibfnamefont
  {F.}~\bibnamefont {Kuemmeth}}, \bibinfo {author} {\bibfnamefont
  {H.}~\bibnamefont {Bluhm}}, \bibinfo {author} {\bibfnamefont
  {J.}~\bibnamefont {Pla}}, \bibinfo {author} {\bibfnamefont {C.}~\bibnamefont
  {Hill}}, \bibinfo {author} {\bibfnamefont {J.}~\bibnamefont {Salfi}},
  \bibinfo {author} {\bibfnamefont {A.}~\bibnamefont {Oiwa}}, \bibinfo {author}
  {\bibfnamefont {J.~T.}\ \bibnamefont {Muhonen}}, \bibinfo {author}
  {\bibfnamefont {E.}~\bibnamefont {Verhagen}}, \bibinfo {author}
  {\bibfnamefont {M.~D.}\ \bibnamefont {LaHaye}}, \bibinfo {author}
  {\bibfnamefont {H.~H.}\ \bibnamefont {Kim}}, \bibinfo {author} {\bibfnamefont
  {A.~W.}\ \bibnamefont {Tsen}}, \bibinfo {author} {\bibfnamefont
  {D.}~\bibnamefont {Culcer}}, \bibinfo {author} {\bibfnamefont
  {A.}~\bibnamefont {Geresdi}}, \bibinfo {author} {\bibfnamefont {J.~A.}\
  \bibnamefont {Mol}}, \bibinfo {author} {\bibfnamefont {V.}~\bibnamefont
  {Mohan}}, \bibinfo {author} {\bibfnamefont {P.~K.}\ \bibnamefont {Jain}},\
  and\ \bibinfo {author} {\bibfnamefont {J.}~\bibnamefont {Baugh}},\ }\bibfield
   {title} {\enquote {\bibinfo {title} {Roadmap on quantum nanotechnologies},}\
  }\href {https://doi.org/10.1088/1361-6528/abb333} {\bibfield  {journal}
  {\bibinfo  {journal} {Nanotechnology}\ }\textbf {\bibinfo {volume} {32}},\
  \bibinfo {pages} {162003} (\bibinfo {year} {2021})}\BibitemShut {NoStop}%
\bibitem [{\citenamefont {Andrianov}\ and\ \citenamefont
  {Moiseev}(2014)}]{Andrianov2014}%
  \BibitemOpen
  \bibfield  {author} {\bibinfo {author} {\bibfnamefont {S.~N.}\ \bibnamefont
  {Andrianov}}\ and\ \bibinfo {author} {\bibfnamefont {S.~A.}\ \bibnamefont
  {Moiseev}},\ }\bibfield  {title} {\enquote {\bibinfo {title} {Magnon qubit
  and quantum computing on magnon bose-einstein condensates},}\ }\href
  {https://doi.org/10.1103/PhysRevA.90.042303} {\bibfield  {journal} {\bibinfo
  {journal} {Phys. Rev. A}\ }\textbf {\bibinfo {volume} {90}},\ \bibinfo
  {pages} {042303} (\bibinfo {year} {2014})}\BibitemShut {NoStop}%
\bibitem [{\citenamefont {Het\'enyi}\ \emph {et~al.}(2022)\citenamefont
  {Het\'enyi}, \citenamefont {Mook}, \citenamefont {Klinovaja},\ and\
  \citenamefont {Loss}}]{Hetenyi2022}%
  \BibitemOpen
  \bibfield  {author} {\bibinfo {author} {\bibfnamefont {B.}~\bibnamefont
  {Het\'enyi}}, \bibinfo {author} {\bibfnamefont {A.}~\bibnamefont {Mook}},
  \bibinfo {author} {\bibfnamefont {J.}~\bibnamefont {Klinovaja}},\ and\
  \bibinfo {author} {\bibfnamefont {D.}~\bibnamefont {Loss}},\ }\bibfield
  {title} {\enquote {\bibinfo {title} {Long-distance coupling of spin qubits
  via topological magnons},}\ }\href
  {https://doi.org/10.1103/PhysRevB.106.235409} {\bibfield  {journal} {\bibinfo
   {journal} {Phys. Rev. B}\ }\textbf {\bibinfo {volume} {106}},\ \bibinfo
  {pages} {235409} (\bibinfo {year} {2022})}\BibitemShut {NoStop}%
\bibitem [{\citenamefont {Terhal}, \citenamefont {Conrad},\ and\ \citenamefont
  {Vuillot}(2020)}]{Terhal2020}%
  \BibitemOpen
  \bibfield  {author} {\bibinfo {author} {\bibfnamefont {B.~M.}\ \bibnamefont
  {Terhal}}, \bibinfo {author} {\bibfnamefont {J.}~\bibnamefont {Conrad}},\
  and\ \bibinfo {author} {\bibfnamefont {C.}~\bibnamefont {Vuillot}},\
  }\bibfield  {title} {\enquote {\bibinfo {title} {Towards scalable bosonic
  quantum error correction},}\ }\href
  {https://doi.org/10.1088/2058-9565/ab98a5} {\bibfield  {journal} {\bibinfo
  {journal} {Quantum Science and Technology}\ }\textbf {\bibinfo {volume}
  {5}},\ \bibinfo {pages} {043001} (\bibinfo {year} {2020})}\BibitemShut
  {NoStop}%
\bibitem [{\citenamefont {Bejarano}\ \emph {et~al.}(2024)\citenamefont
  {Bejarano}, \citenamefont {Goncalves}, \citenamefont {Hache}, \citenamefont
  {Hollenbach}, \citenamefont {Heins}, \citenamefont {Hula}, \citenamefont
  {Körber}, \citenamefont {Heinze}, \citenamefont {Berencén}, \citenamefont
  {Helm}, \citenamefont {Fassbender}, \citenamefont {Astakhov},\ and\
  \citenamefont {Schultheiss}}]{Bejarano2024}%
  \BibitemOpen
  \bibfield  {author} {\bibinfo {author} {\bibfnamefont {M.}~\bibnamefont
  {Bejarano}}, \bibinfo {author} {\bibfnamefont {F.~J.~T.}\ \bibnamefont
  {Goncalves}}, \bibinfo {author} {\bibfnamefont {T.}~\bibnamefont {Hache}},
  \bibinfo {author} {\bibfnamefont {M.}~\bibnamefont {Hollenbach}}, \bibinfo
  {author} {\bibfnamefont {C.}~\bibnamefont {Heins}}, \bibinfo {author}
  {\bibfnamefont {T.}~\bibnamefont {Hula}}, \bibinfo {author} {\bibfnamefont
  {L.}~\bibnamefont {Körber}}, \bibinfo {author} {\bibfnamefont
  {J.}~\bibnamefont {Heinze}}, \bibinfo {author} {\bibfnamefont
  {Y.}~\bibnamefont {Berencén}}, \bibinfo {author} {\bibfnamefont
  {M.}~\bibnamefont {Helm}}, \bibinfo {author} {\bibfnamefont {J.}~\bibnamefont
  {Fassbender}}, \bibinfo {author} {\bibfnamefont {G.~V.}\ \bibnamefont
  {Astakhov}},\ and\ \bibinfo {author} {\bibfnamefont {H.}~\bibnamefont
  {Schultheiss}},\ }\bibfield  {title} {\enquote {\bibinfo {title} {Parametric
  magnon transduction to spin qubits},}\ }\href
  {https://doi.org/10.1126/sciadv.adi2042} {\bibfield  {journal} {\bibinfo
  {journal} {Science Advances}\ }\textbf {\bibinfo {volume} {10}},\ \bibinfo
  {pages} {eadi2042} (\bibinfo {year} {2024})}\BibitemShut {NoStop}%
\bibitem [{\citenamefont {Walls}(1983)}]{Walls1983}%
  \BibitemOpen
  \bibfield  {author} {\bibinfo {author} {\bibfnamefont {D.~F.}\ \bibnamefont
  {Walls}},\ }\bibfield  {title} {\enquote {\bibinfo {title} {Squeezed states
  of light},}\ }\href {https://doi.org/10.1038/306141a0} {\bibfield  {journal}
  {\bibinfo  {journal} {Nature}\ }\textbf {\bibinfo {volume} {306}},\ \bibinfo
  {pages} {141--146} (\bibinfo {year} {1983})}\BibitemShut {NoStop}%
\bibitem [{\citenamefont {Slusher}\ \emph {et~al.}(1985)\citenamefont
  {Slusher}, \citenamefont {Hollberg}, \citenamefont {Yurke}, \citenamefont
  {Mertz},\ and\ \citenamefont {Valley}}]{Slusher1985}%
  \BibitemOpen
  \bibfield  {author} {\bibinfo {author} {\bibfnamefont {R.~E.}\ \bibnamefont
  {Slusher}}, \bibinfo {author} {\bibfnamefont {L.~W.}\ \bibnamefont
  {Hollberg}}, \bibinfo {author} {\bibfnamefont {B.}~\bibnamefont {Yurke}},
  \bibinfo {author} {\bibfnamefont {J.~C.}\ \bibnamefont {Mertz}},\ and\
  \bibinfo {author} {\bibfnamefont {J.~F.}\ \bibnamefont {Valley}},\ }\bibfield
   {title} {\enquote {\bibinfo {title} {Observation of squeezed states
  generated by four-wave mixing in an optical cavity},}\ }\href
  {https://doi.org/10.1103/PhysRevLett.55.2409} {\bibfield  {journal} {\bibinfo
   {journal} {Phys. Rev. Lett.}\ }\textbf {\bibinfo {volume} {55}},\ \bibinfo
  {pages} {2409--2412} (\bibinfo {year} {1985})}\BibitemShut {NoStop}%
\bibitem [{\citenamefont {Wu}\ \emph {et~al.}(1986)\citenamefont {Wu},
  \citenamefont {Kimble}, \citenamefont {Hall},\ and\ \citenamefont
  {Wu}}]{Wu1986}%
  \BibitemOpen
  \bibfield  {author} {\bibinfo {author} {\bibfnamefont {L.-A.}\ \bibnamefont
  {Wu}}, \bibinfo {author} {\bibfnamefont {H.~J.}\ \bibnamefont {Kimble}},
  \bibinfo {author} {\bibfnamefont {J.~L.}\ \bibnamefont {Hall}},\ and\
  \bibinfo {author} {\bibfnamefont {H.}~\bibnamefont {Wu}},\ }\bibfield
  {title} {\enquote {\bibinfo {title} {Generation of squeezed states by
  parametric down conversion},}\ }\href
  {https://doi.org/10.1103/PhysRevLett.57.2520} {\bibfield  {journal} {\bibinfo
   {journal} {Phys. Rev. Lett.}\ }\textbf {\bibinfo {volume} {57}},\ \bibinfo
  {pages} {2520--2523} (\bibinfo {year} {1986})}\BibitemShut {NoStop}%
\bibitem [{\citenamefont {Collaboration}(2011)}]{Abadie2011}%
  \BibitemOpen
  \bibfield  {author} {\bibinfo {author} {\bibfnamefont {T.~L.~S.}\
  \bibnamefont {Collaboration}},\ }\bibfield  {title} {\enquote {\bibinfo
  {title} {A gravitational wave observatory operating beyond the quantum
  shot-noise limit},}\ }\href {https://doi.org/10.1038/nphys2083} {\bibfield
  {journal} {\bibinfo  {journal} {Nature Physics}\ }\textbf {\bibinfo {volume}
  {7}},\ \bibinfo {pages} {962--965} (\bibinfo {year} {2011})}\BibitemShut
  {NoStop}%
\bibitem [{\citenamefont {Collaboration}(2013)}]{Aasi2013}%
  \BibitemOpen
  \bibfield  {author} {\bibinfo {author} {\bibfnamefont {T.~L.~S.}\
  \bibnamefont {Collaboration}},\ }\bibfield  {title} {\enquote {\bibinfo
  {title} {Enhanced sensitivity of the ligo gravitational wave detector by
  using squeezed states of light},}\ }\href
  {https://doi.org/10.1038/nphoton.2013.177} {\bibfield  {journal} {\bibinfo
  {journal} {Nature Photonics}\ }\textbf {\bibinfo {volume} {7}},\ \bibinfo
  {pages} {613--619} (\bibinfo {year} {2013})}\BibitemShut {NoStop}%
\bibitem [{\citenamefont {Li}, \citenamefont {Zhu},\ and\ \citenamefont
  {Agarwal}(2019)}]{Li2019}%
  \BibitemOpen
  \bibfield  {author} {\bibinfo {author} {\bibfnamefont {J.}~\bibnamefont
  {Li}}, \bibinfo {author} {\bibfnamefont {S.-Y.}\ \bibnamefont {Zhu}},\ and\
  \bibinfo {author} {\bibfnamefont {G.~S.}\ \bibnamefont {Agarwal}},\
  }\bibfield  {title} {\enquote {\bibinfo {title} {Squeezed states of magnons
  and phonons in cavity magnomechanics},}\ }\href
  {https://doi.org/10.1103/PhysRevA.99.021801} {\bibfield  {journal} {\bibinfo
  {journal} {Phys. Rev. A}\ }\textbf {\bibinfo {volume} {99}},\ \bibinfo
  {pages} {021801} (\bibinfo {year} {2019})}\BibitemShut {NoStop}%
\bibitem [{\citenamefont {Hioki}\ \emph {et~al.}(2026)\citenamefont {Hioki},
  \citenamefont {Tojo}, \citenamefont {Elyasi}, \citenamefont {Horibe},
  \citenamefont {Shimizu}, \citenamefont {Hoshi}, \citenamefont {Makiuchi},
  \citenamefont {Bauer},\ and\ \citenamefont {Saitoh}}]{Hioki2026}%
  \BibitemOpen
  \bibfield  {author} {\bibinfo {author} {\bibfnamefont {T.}~\bibnamefont
  {Hioki}}, \bibinfo {author} {\bibfnamefont {K.}~\bibnamefont {Tojo}},
  \bibinfo {author} {\bibfnamefont {M.}~\bibnamefont {Elyasi}}, \bibinfo
  {author} {\bibfnamefont {S.}~\bibnamefont {Horibe}}, \bibinfo {author}
  {\bibfnamefont {H.}~\bibnamefont {Shimizu}}, \bibinfo {author} {\bibfnamefont
  {K.}~\bibnamefont {Hoshi}}, \bibinfo {author} {\bibfnamefont
  {T.}~\bibnamefont {Makiuchi}}, \bibinfo {author} {\bibfnamefont {G.~E.~W.}\
  \bibnamefont {Bauer}},\ and\ \bibinfo {author} {\bibfnamefont
  {E.}~\bibnamefont {Saitoh}},\ }\bibfield  {title} {\enquote {\bibinfo {title}
  {Single- and two-mode magnon thermal squeezing},}\ }\href
  {https://doi.org/10.1038/s41567-026-03294-4} {\bibfield  {journal} {\bibinfo
  {journal} {Nature Physics}\ } (\bibinfo {year} {2026}),\
  10.1038/s41567-026-03294-4}\BibitemShut {NoStop}%
\bibitem [{\citenamefont {Guo}\ \emph {et~al.}(2023)\citenamefont {Guo},
  \citenamefont {Cheng}, \citenamefont {Tan},\ and\ \citenamefont
  {Li}}]{Guo2023}%
  \BibitemOpen
  \bibfield  {author} {\bibinfo {author} {\bibfnamefont {Q.}~\bibnamefont
  {Guo}}, \bibinfo {author} {\bibfnamefont {J.}~\bibnamefont {Cheng}}, \bibinfo
  {author} {\bibfnamefont {H.}~\bibnamefont {Tan}},\ and\ \bibinfo {author}
  {\bibfnamefont {J.}~\bibnamefont {Li}},\ }\bibfield  {title} {\enquote
  {\bibinfo {title} {Magnon squeezing by two-tone driving of a qubit in
  cavity-magnon-qubit systems},}\ }\href
  {https://doi.org/10.1103/PhysRevA.108.063703} {\bibfield  {journal} {\bibinfo
   {journal} {Phys. Rev. A}\ }\textbf {\bibinfo {volume} {108}},\ \bibinfo
  {pages} {063703} (\bibinfo {year} {2023})}\BibitemShut {NoStop}%
\bibitem [{\citenamefont {Kamra}\ and\ \citenamefont
  {Belzig}(2016{\natexlab{b}})}]{Kamra2016_hybrid}%
  \BibitemOpen
  \bibfield  {author} {\bibinfo {author} {\bibfnamefont {A.}~\bibnamefont
  {Kamra}}\ and\ \bibinfo {author} {\bibfnamefont {W.}~\bibnamefont {Belzig}},\
  }\bibfield  {title} {\enquote {\bibinfo {title} {Magnon-mediated spin current
  noise in ferromagnet $|$ nonmagnetic conductor hybrids},}\ }\href
  {https://doi.org/10.1103/PhysRevB.94.014419} {\bibfield  {journal} {\bibinfo
  {journal} {Phys. Rev. B}\ }\textbf {\bibinfo {volume} {94}},\ \bibinfo
  {pages} {014419} (\bibinfo {year} {2016}{\natexlab{b}})}\BibitemShut
  {NoStop}%
\bibitem [{\citenamefont {Kamra}, \citenamefont {Belzig},\ and\ \citenamefont
  {Brataas}(2020)}]{Kamra2020}%
  \BibitemOpen
  \bibfield  {author} {\bibinfo {author} {\bibfnamefont {A.}~\bibnamefont
  {Kamra}}, \bibinfo {author} {\bibfnamefont {W.}~\bibnamefont {Belzig}},\ and\
  \bibinfo {author} {\bibfnamefont {A.}~\bibnamefont {Brataas}},\ }\bibfield
  {title} {\enquote {\bibinfo {title} {Magnon-squeezing as a niche of quantum
  magnonics},}\ }\href {https://doi.org/10.1063/5.0021099} {\bibfield
  {journal} {\bibinfo  {journal} {Applied Physics Letters}\ }\textbf {\bibinfo
  {volume} {117}},\ \bibinfo {pages} {090501} (\bibinfo {year}
  {2020})}\BibitemShut {NoStop}%
\bibitem [{\citenamefont {R\"omling}\ \emph {et~al.}(2023)\citenamefont
  {R\"omling}, \citenamefont {Vivas-Via\~na}, \citenamefont {Mu\~noz},\ and\
  \citenamefont {Kamra}}]{Roemling2023}%
  \BibitemOpen
  \bibfield  {author} {\bibinfo {author} {\bibfnamefont {A.-L.~E.}\
  \bibnamefont {R\"omling}}, \bibinfo {author} {\bibfnamefont {A.}~\bibnamefont
  {Vivas-Via\~na}}, \bibinfo {author} {\bibfnamefont {C.~S.}\ \bibnamefont
  {Mu\~noz}},\ and\ \bibinfo {author} {\bibfnamefont {A.}~\bibnamefont
  {Kamra}},\ }\bibfield  {title} {\enquote {\bibinfo {title} {Resolving
  nonclassical magnon composition of a magnetic ground state via a qubit},}\
  }\href {https://doi.org/10.1103/PhysRevLett.131.143602} {\bibfield  {journal}
  {\bibinfo  {journal} {Phys. Rev. Lett.}\ }\textbf {\bibinfo {volume} {131}},\
  \bibinfo {pages} {143602} (\bibinfo {year} {2023})}\BibitemShut {NoStop}%
\bibitem [{\citenamefont {Zou}, \citenamefont {Kim},\ and\ \citenamefont
  {Tserkovnyak}(2020)}]{Zou2020}%
  \BibitemOpen
  \bibfield  {author} {\bibinfo {author} {\bibfnamefont {J.}~\bibnamefont
  {Zou}}, \bibinfo {author} {\bibfnamefont {S.~K.}\ \bibnamefont {Kim}},\ and\
  \bibinfo {author} {\bibfnamefont {Y.}~\bibnamefont {Tserkovnyak}},\
  }\bibfield  {title} {\enquote {\bibinfo {title} {Tuning entanglement by
  squeezing magnons in anisotropic magnets},}\ }\href
  {https://doi.org/10.1103/PhysRevB.101.014416} {\bibfield  {journal} {\bibinfo
   {journal} {Phys. Rev. B}\ }\textbf {\bibinfo {volume} {101}},\ \bibinfo
  {pages} {014416} (\bibinfo {year} {2020})}\BibitemShut {NoStop}%
\bibitem [{\citenamefont {Elyasi}, \citenamefont {Blanter},\ and\ \citenamefont
  {Bauer}(2020)}]{Elyasi2020}%
  \BibitemOpen
  \bibfield  {author} {\bibinfo {author} {\bibfnamefont {M.}~\bibnamefont
  {Elyasi}}, \bibinfo {author} {\bibfnamefont {Y.~M.}\ \bibnamefont
  {Blanter}},\ and\ \bibinfo {author} {\bibfnamefont {G.~E.~W.}\ \bibnamefont
  {Bauer}},\ }\bibfield  {title} {\enquote {\bibinfo {title} {Resources of
  nonlinear cavity magnonics for quantum information},}\ }\href
  {https://doi.org/10.1103/PhysRevB.101.054402} {\bibfield  {journal} {\bibinfo
   {journal} {Phys. Rev. B}\ }\textbf {\bibinfo {volume} {101}},\ \bibinfo
  {pages} {054402} (\bibinfo {year} {2020})}\BibitemShut {NoStop}%
\bibitem [{\citenamefont {Degen}, \citenamefont {Reinhard},\ and\ \citenamefont
  {Cappellaro}(2017)}]{Degen2017}%
  \BibitemOpen
  \bibfield  {author} {\bibinfo {author} {\bibfnamefont {C.~L.}\ \bibnamefont
  {Degen}}, \bibinfo {author} {\bibfnamefont {F.}~\bibnamefont {Reinhard}},\
  and\ \bibinfo {author} {\bibfnamefont {P.}~\bibnamefont {Cappellaro}},\
  }\bibfield  {title} {\enquote {\bibinfo {title} {Quantum sensing},}\ }\href
  {https://doi.org/10.1103/RevModPhys.89.035002} {\bibfield  {journal}
  {\bibinfo  {journal} {Rev. Mod. Phys.}\ }\textbf {\bibinfo {volume} {89}},\
  \bibinfo {pages} {035002} (\bibinfo {year} {2017})}\BibitemShut {NoStop}%
\bibitem [{\citenamefont {Patel}\ and\ \citenamefont
  {Desai}(2025)}]{PatelDesai2025}%
  \BibitemOpen
  \bibfield  {author} {\bibinfo {author} {\bibfnamefont {P.~S.}\ \bibnamefont
  {Patel}}\ and\ \bibinfo {author} {\bibfnamefont {D.~B.}\ \bibnamefont
  {Desai}},\ }\bibfield  {title} {\enquote {\bibinfo {title} {Review of
  qubit-based quantum sensing},}\ }\href
  {https://doi.org/10.1007/s11128-025-04699-5} {\bibfield  {journal} {\bibinfo
  {journal} {Quantum Information Processing}\ }\textbf {\bibinfo {volume}
  {24}},\ \bibinfo {pages} {83} (\bibinfo {year} {2025})}\BibitemShut {NoStop}%
\bibitem [{\citenamefont {Fink}\ \emph {et~al.}(2024)\citenamefont {Fink},
  \citenamefont {Salemi}, \citenamefont {Young}, \citenamefont {Schuster},\
  and\ \citenamefont {Kurinsky}}]{Fink2024}%
  \BibitemOpen
  \bibfield  {author} {\bibinfo {author} {\bibfnamefont {C.}~\bibnamefont
  {Fink}}, \bibinfo {author} {\bibfnamefont {C.}~\bibnamefont {Salemi}},
  \bibinfo {author} {\bibfnamefont {B.}~\bibnamefont {Young}}, \bibinfo
  {author} {\bibfnamefont {D.}~\bibnamefont {Schuster}},\ and\ \bibinfo
  {author} {\bibfnamefont {N.}~\bibnamefont {Kurinsky}},\ }\bibfield  {title}
  {\enquote {\bibinfo {title} {Superconducting quasiparticle-amplifying
  transmon: A qubit-based sensor for mev-scale phonons and single terahertz
  photons},}\ }\href {https://doi.org/10.1103/PhysRevApplied.22.054009}
  {\bibfield  {journal} {\bibinfo  {journal} {Phys. Rev. Appl.}\ }\textbf
  {\bibinfo {volume} {22}},\ \bibinfo {pages} {054009} (\bibinfo {year}
  {2024})}\BibitemShut {NoStop}%
\bibitem [{\citenamefont {Kakuyanagi}\ \emph {et~al.}(2023)\citenamefont
  {Kakuyanagi}, \citenamefont {Toida}, \citenamefont {Abdurakhimov},\ and\
  \citenamefont {Saito}}]{Kakuyanagi2023}%
  \BibitemOpen
  \bibfield  {author} {\bibinfo {author} {\bibfnamefont {K.}~\bibnamefont
  {Kakuyanagi}}, \bibinfo {author} {\bibfnamefont {H.}~\bibnamefont {Toida}},
  \bibinfo {author} {\bibfnamefont {L.~V.}\ \bibnamefont {Abdurakhimov}},\ and\
  \bibinfo {author} {\bibfnamefont {S.}~\bibnamefont {Saito}},\ }\bibfield
  {title} {\enquote {\bibinfo {title} {Submicrometer‑scale temperature
  sensing using quantum coherence of a superconducting qubit},}\ }\href
  {https://doi.org/10.1088/1367-2630/acb379} {\bibfield  {journal} {\bibinfo
  {journal} {New Journal of Physics}\ }\textbf {\bibinfo {volume} {25}},\
  \bibinfo {pages} {013036} (\bibinfo {year} {2023})}\BibitemShut {NoStop}%
\bibitem [{\citenamefont {Danilin}\ and\ \citenamefont
  {Weides}(2021)}]{DanilinWeides2021}%
  \BibitemOpen
  \bibfield  {author} {\bibinfo {author} {\bibfnamefont {S.}~\bibnamefont
  {Danilin}}\ and\ \bibinfo {author} {\bibfnamefont {M.}~\bibnamefont
  {Weides}},\ }\href {https://arxiv.org/abs/2103.11022} {\enquote {\bibinfo
  {title} {Quantum sensing with superconducting circuits},}\ }\bibinfo
  {howpublished} {arXiv preprint arXiv:2103.11022} (\bibinfo {year}
  {2021})\BibitemShut {NoStop}%
\bibitem [{\citenamefont {Casola}, \citenamefont {van~der Sar},\ and\
  \citenamefont {Yacoby}(2018)}]{Casola2018}%
  \BibitemOpen
  \bibfield  {author} {\bibinfo {author} {\bibfnamefont {F.}~\bibnamefont
  {Casola}}, \bibinfo {author} {\bibfnamefont {T.}~\bibnamefont {van~der
  Sar}},\ and\ \bibinfo {author} {\bibfnamefont {A.}~\bibnamefont {Yacoby}},\
  }\bibfield  {title} {\enquote {\bibinfo {title} {Probing condensed matter
  physics with magnetometry based on nitrogen-vacancy centres in diamond},}\
  }\href {https://doi.org/10.1038/natrevmats.2017.88} {\bibfield  {journal}
  {\bibinfo  {journal} {Nature Reviews Materials}\ }\textbf {\bibinfo {volume}
  {3}},\ \bibinfo {pages} {17088} (\bibinfo {year} {2018})}\BibitemShut
  {NoStop}%
\bibitem [{\citenamefont {Xu}, \citenamefont {Zhang},\ and\ \citenamefont
  {Tian}(2023)}]{Xu2023}%
  \BibitemOpen
  \bibfield  {author} {\bibinfo {author} {\bibfnamefont {Y.}~\bibnamefont
  {Xu}}, \bibinfo {author} {\bibfnamefont {W.}~\bibnamefont {Zhang}},\ and\
  \bibinfo {author} {\bibfnamefont {C.}~\bibnamefont {Tian}},\ }\bibfield
  {title} {\enquote {\bibinfo {title} {Recent advances on applications of nv
  centers; magnetometry in condensed matter physics},}\ }\href
  {https://doi.org/10.1364/PRJ.471266} {\bibfield  {journal} {\bibinfo
  {journal} {Photon. Res.}\ }\textbf {\bibinfo {volume} {11}},\ \bibinfo
  {pages} {393--412} (\bibinfo {year} {2023})}\BibitemShut {NoStop}%
\bibitem [{\citenamefont {Simon}\ \emph {et~al.}(2022)\citenamefont {Simon},
  \citenamefont {Kurdi}, \citenamefont {Carmiggelt}, \citenamefont {Borst},
  \citenamefont {Katan},\ and\ \citenamefont {van~der Sar}}]{Simon2022}%
  \BibitemOpen
  \bibfield  {author} {\bibinfo {author} {\bibfnamefont {B.~G.}\ \bibnamefont
  {Simon}}, \bibinfo {author} {\bibfnamefont {S.}~\bibnamefont {Kurdi}},
  \bibinfo {author} {\bibfnamefont {J.~J.}\ \bibnamefont {Carmiggelt}},
  \bibinfo {author} {\bibfnamefont {M.}~\bibnamefont {Borst}}, \bibinfo
  {author} {\bibfnamefont {A.~J.}\ \bibnamefont {Katan}},\ and\ \bibinfo
  {author} {\bibfnamefont {T.}~\bibnamefont {van~der Sar}},\ }\bibfield
  {title} {\enquote {\bibinfo {title} {Filtering and imaging of
  frequency-degenerate spin waves using nanopositioning of a single-spin
  sensor},}\ }\href {https://doi.org/10.1021/acs.nanolett.2c02791} {\bibfield
  {journal} {\bibinfo  {journal} {Nano Letters}\ }\textbf {\bibinfo {volume}
  {22}},\ \bibinfo {pages} {9198--9204} (\bibinfo {year} {2022})}\BibitemShut
  {NoStop}%
\bibitem [{\citenamefont {van~der Sar}\ \emph {et~al.}(2015)\citenamefont
  {van~der Sar}, \citenamefont {Casola}, \citenamefont {Walsworth},\ and\
  \citenamefont {Yacoby}}]{Sar2015}%
  \BibitemOpen
  \bibfield  {author} {\bibinfo {author} {\bibfnamefont {T.}~\bibnamefont
  {van~der Sar}}, \bibinfo {author} {\bibfnamefont {F.}~\bibnamefont {Casola}},
  \bibinfo {author} {\bibfnamefont {R.}~\bibnamefont {Walsworth}},\ and\
  \bibinfo {author} {\bibfnamefont {A.}~\bibnamefont {Yacoby}},\ }\bibfield
  {title} {\enquote {\bibinfo {title} {Nanometre-scale probing of spin waves
  using single electron spins},}\ }\href {https://doi.org/10.1038/ncomms8886}
  {\bibfield  {journal} {\bibinfo  {journal} {Nature Communications}\ }\textbf
  {\bibinfo {volume} {6}},\ \bibinfo {pages} {7886} (\bibinfo {year}
  {2015})}\BibitemShut {NoStop}%
\bibitem [{\citenamefont {Page}\ \emph {et~al.}(2019)\citenamefont {Page},
  \citenamefont {McCullian}, \citenamefont {Purser}, \citenamefont {Schulze},
  \citenamefont {Nakatani}, \citenamefont {Wolfe}, \citenamefont {Childress},
  \citenamefont {McConney}, \citenamefont {Howe}, \citenamefont {Hammel},\ and\
  \citenamefont {Bhallamudi}}]{Page_2019}%
  \BibitemOpen
  \bibfield  {author} {\bibinfo {author} {\bibfnamefont {M.~R.}\ \bibnamefont
  {Page}}, \bibinfo {author} {\bibfnamefont {B.~A.}\ \bibnamefont {McCullian}},
  \bibinfo {author} {\bibfnamefont {C.~M.}\ \bibnamefont {Purser}}, \bibinfo
  {author} {\bibfnamefont {J.~G.}\ \bibnamefont {Schulze}}, \bibinfo {author}
  {\bibfnamefont {T.~M.}\ \bibnamefont {Nakatani}}, \bibinfo {author}
  {\bibfnamefont {C.~S.}\ \bibnamefont {Wolfe}}, \bibinfo {author}
  {\bibfnamefont {J.~R.}\ \bibnamefont {Childress}}, \bibinfo {author}
  {\bibfnamefont {M.~E.}\ \bibnamefont {McConney}}, \bibinfo {author}
  {\bibfnamefont {B.~M.}\ \bibnamefont {Howe}}, \bibinfo {author}
  {\bibfnamefont {P.~C.}\ \bibnamefont {Hammel}},\ and\ \bibinfo {author}
  {\bibfnamefont {V.~P.}\ \bibnamefont {Bhallamudi}},\ }\bibfield  {title}
  {\enquote {\bibinfo {title} {Optically detected ferromagnetic resonance in
  diverse ferromagnets via nitrogen vacancy centers in diamond},}\ }\href
  {https://doi.org/10.1063/1.5083991} {\bibfield  {journal} {\bibinfo
  {journal} {Journal of Applied Physics}\ }\textbf {\bibinfo {volume} {126}},\
  \bibinfo {pages} {124902} (\bibinfo {year} {2019})}\BibitemShut {NoStop}%
\bibitem [{\citenamefont {Dolgirev}\ \emph {et~al.}(2022)\citenamefont
  {Dolgirev}, \citenamefont {Chatterjee}, \citenamefont {Esterlis},
  \citenamefont {Zibrov}, \citenamefont {Lukin}, \citenamefont {Yao},\ and\
  \citenamefont {Demler}}]{Dolgirev2022}%
  \BibitemOpen
  \bibfield  {author} {\bibinfo {author} {\bibfnamefont {P.~E.}\ \bibnamefont
  {Dolgirev}}, \bibinfo {author} {\bibfnamefont {S.}~\bibnamefont
  {Chatterjee}}, \bibinfo {author} {\bibfnamefont {I.}~\bibnamefont
  {Esterlis}}, \bibinfo {author} {\bibfnamefont {A.~A.}\ \bibnamefont
  {Zibrov}}, \bibinfo {author} {\bibfnamefont {M.~D.}\ \bibnamefont {Lukin}},
  \bibinfo {author} {\bibfnamefont {N.~Y.}\ \bibnamefont {Yao}},\ and\ \bibinfo
  {author} {\bibfnamefont {E.}~\bibnamefont {Demler}},\ }\bibfield  {title}
  {\enquote {\bibinfo {title} {Characterizing two-dimensional superconductivity
  via nanoscale noise magnetometry with single-spin qubits},}\ }\href
  {https://doi.org/10.1103/PhysRevB.105.024507} {\bibfield  {journal} {\bibinfo
   {journal} {Phys. Rev. B}\ }\textbf {\bibinfo {volume} {105}},\ \bibinfo
  {pages} {024507} (\bibinfo {year} {2022})}\BibitemShut {NoStop}%
\bibitem [{\citenamefont {Chatterjee}\ \emph {et~al.}(2022)\citenamefont
  {Chatterjee}, \citenamefont {Dolgirev}, \citenamefont {Esterlis},
  \citenamefont {Zibrov}, \citenamefont {Lukin}, \citenamefont {Yao},\ and\
  \citenamefont {Demler}}]{Chatterjee2022}%
  \BibitemOpen
  \bibfield  {author} {\bibinfo {author} {\bibfnamefont {S.}~\bibnamefont
  {Chatterjee}}, \bibinfo {author} {\bibfnamefont {P.~E.}\ \bibnamefont
  {Dolgirev}}, \bibinfo {author} {\bibfnamefont {I.}~\bibnamefont {Esterlis}},
  \bibinfo {author} {\bibfnamefont {A.~A.}\ \bibnamefont {Zibrov}}, \bibinfo
  {author} {\bibfnamefont {M.~D.}\ \bibnamefont {Lukin}}, \bibinfo {author}
  {\bibfnamefont {N.~Y.}\ \bibnamefont {Yao}},\ and\ \bibinfo {author}
  {\bibfnamefont {E.}~\bibnamefont {Demler}},\ }\bibfield  {title} {\enquote
  {\bibinfo {title} {Single-spin qubit magnetic spectroscopy of two-dimensional
  superconductivity},}\ }\href
  {https://doi.org/10.1103/PhysRevResearch.4.L012001} {\bibfield  {journal}
  {\bibinfo  {journal} {Phys. Rev. Res.}\ }\textbf {\bibinfo {volume} {4}},\
  \bibinfo {pages} {L012001} (\bibinfo {year} {2022})}\BibitemShut {NoStop}%
\bibitem [{\citenamefont {Bhattacharyya}\ \emph {et~al.}(2024)\citenamefont
  {Bhattacharyya}, \citenamefont {Chen}, \citenamefont {Huang}, \citenamefont
  {Chatterjee}, \citenamefont {Huang}, \citenamefont {Kobrin}, \citenamefont
  {Lyu}, \citenamefont {Smart}, \citenamefont {Block}, \citenamefont {Wang},
  \citenamefont {Wang}, \citenamefont {Wu}, \citenamefont {Hsieh},
  \citenamefont {Ma}, \citenamefont {Mandyam}, \citenamefont {Chen},
  \citenamefont {Davis}, \citenamefont {Geballe}, \citenamefont {Zu},
  \citenamefont {Struzhkin}, \citenamefont {Jeanloz}, \citenamefont {Moore},
  \citenamefont {Cui}, \citenamefont {Galli}, \citenamefont {Halperin},
  \citenamefont {Laumann},\ and\ \citenamefont {Yao}}]{Bhattacharyya2024}%
  \BibitemOpen
  \bibfield  {author} {\bibinfo {author} {\bibfnamefont {P.}~\bibnamefont
  {Bhattacharyya}}, \bibinfo {author} {\bibfnamefont {W.}~\bibnamefont {Chen}},
  \bibinfo {author} {\bibfnamefont {X.}~\bibnamefont {Huang}}, \bibinfo
  {author} {\bibfnamefont {S.}~\bibnamefont {Chatterjee}}, \bibinfo {author}
  {\bibfnamefont {B.}~\bibnamefont {Huang}}, \bibinfo {author} {\bibfnamefont
  {B.}~\bibnamefont {Kobrin}}, \bibinfo {author} {\bibfnamefont
  {Y.}~\bibnamefont {Lyu}}, \bibinfo {author} {\bibfnamefont {T.~J.}\
  \bibnamefont {Smart}}, \bibinfo {author} {\bibfnamefont {M.}~\bibnamefont
  {Block}}, \bibinfo {author} {\bibfnamefont {E.}~\bibnamefont {Wang}},
  \bibinfo {author} {\bibfnamefont {Z.}~\bibnamefont {Wang}}, \bibinfo {author}
  {\bibfnamefont {W.}~\bibnamefont {Wu}}, \bibinfo {author} {\bibfnamefont
  {S.}~\bibnamefont {Hsieh}}, \bibinfo {author} {\bibfnamefont
  {H.}~\bibnamefont {Ma}}, \bibinfo {author} {\bibfnamefont {S.}~\bibnamefont
  {Mandyam}}, \bibinfo {author} {\bibfnamefont {B.}~\bibnamefont {Chen}},
  \bibinfo {author} {\bibfnamefont {E.}~\bibnamefont {Davis}}, \bibinfo
  {author} {\bibfnamefont {Z.~M.}\ \bibnamefont {Geballe}}, \bibinfo {author}
  {\bibfnamefont {C.}~\bibnamefont {Zu}}, \bibinfo {author} {\bibfnamefont
  {V.}~\bibnamefont {Struzhkin}}, \bibinfo {author} {\bibfnamefont
  {R.}~\bibnamefont {Jeanloz}}, \bibinfo {author} {\bibfnamefont {J.~E.}\
  \bibnamefont {Moore}}, \bibinfo {author} {\bibfnamefont {T.}~\bibnamefont
  {Cui}}, \bibinfo {author} {\bibfnamefont {G.}~\bibnamefont {Galli}}, \bibinfo
  {author} {\bibfnamefont {B.~I.}\ \bibnamefont {Halperin}}, \bibinfo {author}
  {\bibfnamefont {C.~R.}\ \bibnamefont {Laumann}},\ and\ \bibinfo {author}
  {\bibfnamefont {N.~Y.}\ \bibnamefont {Yao}},\ }\bibfield  {title} {\enquote
  {\bibinfo {title} {Imaging the meissner effect in hydride superconductors
  using quantum sensors},}\ }\href {https://doi.org/10.1038/s41586-024-07026-7}
  {\bibfield  {journal} {\bibinfo  {journal} {Nature}\ }\textbf {\bibinfo
  {volume} {627}},\ \bibinfo {pages} {73--79} (\bibinfo {year}
  {2024})}\BibitemShut {NoStop}%
\bibitem [{\citenamefont {Melendez}\ \emph {et~al.}(2025)\citenamefont
  {Melendez}, \citenamefont {Das}, \citenamefont {Rodriguez}, \citenamefont
  {Kao}, \citenamefont {Liu}, \citenamefont {Williams}, \citenamefont {Lv},
  \citenamefont {Goldberger}, \citenamefont {Chatterjee}, \citenamefont
  {Singh},\ and\ \citenamefont {Hammel}}]{Melendez2025}%
  \BibitemOpen
  \bibfield  {author} {\bibinfo {author} {\bibfnamefont {A.~L.}\ \bibnamefont
  {Melendez}}, \bibinfo {author} {\bibfnamefont {S.}~\bibnamefont {Das}},
  \bibinfo {author} {\bibfnamefont {F.~A.}\ \bibnamefont {Rodriguez}}, \bibinfo
  {author} {\bibfnamefont {I.-H.}\ \bibnamefont {Kao}}, \bibinfo {author}
  {\bibfnamefont {W.}~\bibnamefont {Liu}}, \bibinfo {author} {\bibfnamefont
  {A.~J.}\ \bibnamefont {Williams}}, \bibinfo {author} {\bibfnamefont
  {B.}~\bibnamefont {Lv}}, \bibinfo {author} {\bibfnamefont {J.}~\bibnamefont
  {Goldberger}}, \bibinfo {author} {\bibfnamefont {S.}~\bibnamefont
  {Chatterjee}}, \bibinfo {author} {\bibfnamefont {S.}~\bibnamefont {Singh}},\
  and\ \bibinfo {author} {\bibfnamefont {P.~C.}\ \bibnamefont {Hammel}},\
  }\bibfield  {title} {\enquote {\bibinfo {title} {Quantum sensing of broadband
  spin dynamics and magnon transport in antiferromagnets},}\ }\href
  {https://doi.org/10.1126/sciadv.adu9381} {\bibfield  {journal} {\bibinfo
  {journal} {Science Advances}\ }\textbf {\bibinfo {volume} {11}},\ \bibinfo
  {pages} {eadu9381} (\bibinfo {year} {2025})}\BibitemShut {NoStop}%
\bibitem [{\citenamefont {Machado}\ \emph {et~al.}(2023)\citenamefont
  {Machado}, \citenamefont {Demler}, \citenamefont {Yao},\ and\ \citenamefont
  {Chatterjee}}]{Machado2023}%
  \BibitemOpen
  \bibfield  {author} {\bibinfo {author} {\bibfnamefont {F.}~\bibnamefont
  {Machado}}, \bibinfo {author} {\bibfnamefont {E.~A.}\ \bibnamefont {Demler}},
  \bibinfo {author} {\bibfnamefont {N.~Y.}\ \bibnamefont {Yao}},\ and\ \bibinfo
  {author} {\bibfnamefont {S.}~\bibnamefont {Chatterjee}},\ }\bibfield  {title}
  {\enquote {\bibinfo {title} {Quantum noise spectroscopy of dynamical critical
  phenomena},}\ }\href {https://doi.org/10.1103/PhysRevLett.131.070801}
  {\bibfield  {journal} {\bibinfo  {journal} {Phys. Rev. Lett.}\ }\textbf
  {\bibinfo {volume} {131}},\ \bibinfo {pages} {070801} (\bibinfo {year}
  {2023})}\BibitemShut {NoStop}%
\bibitem [{\citenamefont {Schuster}\ \emph {et~al.}(2007)\citenamefont
  {Schuster}, \citenamefont {Houck}, \citenamefont {Schreier}, \citenamefont
  {Wallraff}, \citenamefont {Gambetta}, \citenamefont {Blais}, \citenamefont
  {Frunzio}, \citenamefont {Majer}, \citenamefont {Johnson}, \citenamefont
  {Devoret}, \citenamefont {Girvin},\ and\ \citenamefont
  {Schoelkopf}}]{Schuster2007}%
  \BibitemOpen
  \bibfield  {author} {\bibinfo {author} {\bibfnamefont {D.~I.}\ \bibnamefont
  {Schuster}}, \bibinfo {author} {\bibfnamefont {A.~A.}\ \bibnamefont {Houck}},
  \bibinfo {author} {\bibfnamefont {J.~A.}\ \bibnamefont {Schreier}}, \bibinfo
  {author} {\bibfnamefont {A.}~\bibnamefont {Wallraff}}, \bibinfo {author}
  {\bibfnamefont {J.~M.}\ \bibnamefont {Gambetta}}, \bibinfo {author}
  {\bibfnamefont {A.}~\bibnamefont {Blais}}, \bibinfo {author} {\bibfnamefont
  {L.}~\bibnamefont {Frunzio}}, \bibinfo {author} {\bibfnamefont
  {J.}~\bibnamefont {Majer}}, \bibinfo {author} {\bibfnamefont
  {B.}~\bibnamefont {Johnson}}, \bibinfo {author} {\bibfnamefont {M.~H.}\
  \bibnamefont {Devoret}}, \bibinfo {author} {\bibfnamefont {S.~M.}\
  \bibnamefont {Girvin}},\ and\ \bibinfo {author} {\bibfnamefont {R.~J.}\
  \bibnamefont {Schoelkopf}},\ }\bibfield  {title} {\enquote {\bibinfo {title}
  {Resolving photon number states in a superconducting circuit},}\ }\href
  {https://doi.org/10.1038/nature05461} {\bibfield  {journal} {\bibinfo
  {journal} {Nature}\ }\textbf {\bibinfo {volume} {445}},\ \bibinfo {pages}
  {515--518} (\bibinfo {year} {2007})}\BibitemShut {NoStop}%
\bibitem [{\citenamefont {Arrangoiz-Arriola}\ \emph {et~al.}(2019)\citenamefont
  {Arrangoiz-Arriola}, \citenamefont {Wollack}, \citenamefont {Wang},
  \citenamefont {Pechal}, \citenamefont {Jiang}, \citenamefont {McKenna},
  \citenamefont {Witmer}, \citenamefont {Van~Laer},\ and\ \citenamefont
  {Safavi-Naeini}}]{Arrangoiz-Arriola2019}%
  \BibitemOpen
  \bibfield  {author} {\bibinfo {author} {\bibfnamefont {P.}~\bibnamefont
  {Arrangoiz-Arriola}}, \bibinfo {author} {\bibfnamefont {E.~A.}\ \bibnamefont
  {Wollack}}, \bibinfo {author} {\bibfnamefont {Z.}~\bibnamefont {Wang}},
  \bibinfo {author} {\bibfnamefont {M.}~\bibnamefont {Pechal}}, \bibinfo
  {author} {\bibfnamefont {W.}~\bibnamefont {Jiang}}, \bibinfo {author}
  {\bibfnamefont {T.~P.}\ \bibnamefont {McKenna}}, \bibinfo {author}
  {\bibfnamefont {J.~D.}\ \bibnamefont {Witmer}}, \bibinfo {author}
  {\bibfnamefont {R.}~\bibnamefont {Van~Laer}},\ and\ \bibinfo {author}
  {\bibfnamefont {A.~H.}\ \bibnamefont {Safavi-Naeini}},\ }\bibfield  {title}
  {\enquote {\bibinfo {title} {Resolving the energy levels of a nanomechanical
  oscillator},}\ }\href {https://doi.org/10.1038/s41586-019-1386-x} {\bibfield
  {journal} {\bibinfo  {journal} {Nature}\ }\textbf {\bibinfo {volume} {571}},\
  \bibinfo {pages} {537--540} (\bibinfo {year} {2019})}\BibitemShut {NoStop}%
\bibitem [{\citenamefont {Peixoto~de Faria}\ and\ \citenamefont
  {Nemes}(1999)}]{Faria1999}%
  \BibitemOpen
  \bibfield  {author} {\bibinfo {author} {\bibfnamefont {J.~G.}\ \bibnamefont
  {Peixoto~de Faria}}\ and\ \bibinfo {author} {\bibfnamefont {M.~C.}\
  \bibnamefont {Nemes}},\ }\bibfield  {title} {\enquote {\bibinfo {title}
  {Dissipative dynamics of the jaynes-cummings model in the dispersive
  approximation: Analytical results},}\ }\href
  {https://doi.org/10.1103/PhysRevA.59.3918} {\bibfield  {journal} {\bibinfo
  {journal} {Phys. Rev. A}\ }\textbf {\bibinfo {volume} {59}},\ \bibinfo
  {pages} {3918--3925} (\bibinfo {year} {1999})}\BibitemShut {NoStop}%
\bibitem [{\citenamefont {Gambetta}\ \emph {et~al.}(2006)\citenamefont
  {Gambetta}, \citenamefont {Blais}, \citenamefont {Schuster}, \citenamefont
  {Wallraff}, \citenamefont {Frunzio}, \citenamefont {Majer}, \citenamefont
  {Devoret}, \citenamefont {Girvin},\ and\ \citenamefont
  {Schoelkopf}}]{Gambetta2006}%
  \BibitemOpen
  \bibfield  {author} {\bibinfo {author} {\bibfnamefont {J.}~\bibnamefont
  {Gambetta}}, \bibinfo {author} {\bibfnamefont {A.}~\bibnamefont {Blais}},
  \bibinfo {author} {\bibfnamefont {D.~I.}\ \bibnamefont {Schuster}}, \bibinfo
  {author} {\bibfnamefont {A.}~\bibnamefont {Wallraff}}, \bibinfo {author}
  {\bibfnamefont {L.}~\bibnamefont {Frunzio}}, \bibinfo {author} {\bibfnamefont
  {J.}~\bibnamefont {Majer}}, \bibinfo {author} {\bibfnamefont {M.~H.}\
  \bibnamefont {Devoret}}, \bibinfo {author} {\bibfnamefont {S.~M.}\
  \bibnamefont {Girvin}},\ and\ \bibinfo {author} {\bibfnamefont {R.~J.}\
  \bibnamefont {Schoelkopf}},\ }\bibfield  {title} {\enquote {\bibinfo {title}
  {Qubit-photon interactions in a cavity: Measurement-induced dephasing and
  number splitting},}\ }\href {https://doi.org/10.1103/PhysRevA.74.042318}
  {\bibfield  {journal} {\bibinfo  {journal} {Phys. Rev. A}\ }\textbf {\bibinfo
  {volume} {74}},\ \bibinfo {pages} {042318} (\bibinfo {year}
  {2006})}\BibitemShut {NoStop}%
\bibitem [{\citenamefont {Xu}\ \emph {et~al.}(2023{\natexlab{a}})\citenamefont
  {Xu}, \citenamefont {Gu}, \citenamefont {Li}, \citenamefont {Weng},
  \citenamefont {Wang}, \citenamefont {Li}, \citenamefont {Wang}, \citenamefont
  {Zhu},\ and\ \citenamefont {You}}]{Wang2023}%
  \BibitemOpen
  \bibfield  {author} {\bibinfo {author} {\bibfnamefont {D.}~\bibnamefont
  {Xu}}, \bibinfo {author} {\bibfnamefont {X.-K.}\ \bibnamefont {Gu}}, \bibinfo
  {author} {\bibfnamefont {H.-K.}\ \bibnamefont {Li}}, \bibinfo {author}
  {\bibfnamefont {Y.-C.}\ \bibnamefont {Weng}}, \bibinfo {author}
  {\bibfnamefont {Y.-P.}\ \bibnamefont {Wang}}, \bibinfo {author}
  {\bibfnamefont {J.}~\bibnamefont {Li}}, \bibinfo {author} {\bibfnamefont
  {H.}~\bibnamefont {Wang}}, \bibinfo {author} {\bibfnamefont {S.-Y.}\
  \bibnamefont {Zhu}},\ and\ \bibinfo {author} {\bibfnamefont {J.~Q.}\
  \bibnamefont {You}},\ }\bibfield  {title} {\enquote {\bibinfo {title}
  {Quantum control of a single magnon in a macroscopic spin system},}\ }\href
  {https://doi.org/10.1103/PhysRevLett.130.193603} {\bibfield  {journal}
  {\bibinfo  {journal} {Phys. Rev. Lett.}\ }\textbf {\bibinfo {volume} {130}},\
  \bibinfo {pages} {193603} (\bibinfo {year} {2023}{\natexlab{a}})}\BibitemShut
  {NoStop}%
\bibitem [{\citenamefont {Kamra}\ \emph {et~al.}(2019)\citenamefont {Kamra},
  \citenamefont {Thingstad}, \citenamefont {Rastelli}, \citenamefont {Duine},
  \citenamefont {Brataas}, \citenamefont {Belzig},\ and\ \citenamefont
  {Sudb\o{}}}]{Kamra2019}%
  \BibitemOpen
  \bibfield  {author} {\bibinfo {author} {\bibfnamefont {A.}~\bibnamefont
  {Kamra}}, \bibinfo {author} {\bibfnamefont {E.}~\bibnamefont {Thingstad}},
  \bibinfo {author} {\bibfnamefont {G.}~\bibnamefont {Rastelli}}, \bibinfo
  {author} {\bibfnamefont {R.~A.}\ \bibnamefont {Duine}}, \bibinfo {author}
  {\bibfnamefont {A.}~\bibnamefont {Brataas}}, \bibinfo {author} {\bibfnamefont
  {W.}~\bibnamefont {Belzig}},\ and\ \bibinfo {author} {\bibfnamefont
  {A.}~\bibnamefont {Sudb\o{}}},\ }\bibfield  {title} {\enquote {\bibinfo
  {title} {Antiferromagnetic magnons as highly squeezed fock states underlying
  quantum correlations},}\ }\href {https://doi.org/10.1103/PhysRevB.100.174407}
  {\bibfield  {journal} {\bibinfo  {journal} {Phys. Rev. B}\ }\textbf {\bibinfo
  {volume} {100}},\ \bibinfo {pages} {174407} (\bibinfo {year}
  {2019})}\BibitemShut {NoStop}%
\bibitem [{\citenamefont {Savary}\ and\ \citenamefont
  {Balents}(2016)}]{Savary2017}%
  \BibitemOpen
  \bibfield  {author} {\bibinfo {author} {\bibfnamefont {L.}~\bibnamefont
  {Savary}}\ and\ \bibinfo {author} {\bibfnamefont {L.}~\bibnamefont
  {Balents}},\ }\bibfield  {title} {\enquote {\bibinfo {title} {Quantum spin
  liquids: a review},}\ }\href {https://doi.org/10.1088/0034-4885/80/1/016502}
  {\bibfield  {journal} {\bibinfo  {journal} {Reports on Progress in Physics}\
  }\textbf {\bibinfo {volume} {80}},\ \bibinfo {pages} {016502} (\bibinfo
  {year} {2016})}\BibitemShut {NoStop}%
\bibitem [{\citenamefont {Zhou}, \citenamefont {Kanoda},\ and\ \citenamefont
  {Ng}(2017)}]{Zhou2017}%
  \BibitemOpen
  \bibfield  {author} {\bibinfo {author} {\bibfnamefont {Y.}~\bibnamefont
  {Zhou}}, \bibinfo {author} {\bibfnamefont {K.}~\bibnamefont {Kanoda}},\ and\
  \bibinfo {author} {\bibfnamefont {T.-K.}\ \bibnamefont {Ng}},\ }\bibfield
  {title} {\enquote {\bibinfo {title} {Quantum spin liquid states},}\ }\href
  {https://doi.org/10.1103/RevModPhys.89.025003} {\bibfield  {journal}
  {\bibinfo  {journal} {Rev. Mod. Phys.}\ }\textbf {\bibinfo {volume} {89}},\
  \bibinfo {pages} {025003} (\bibinfo {year} {2017})}\BibitemShut {NoStop}%
\bibitem [{\citenamefont {Broholm}\ \emph {et~al.}(2020)\citenamefont
  {Broholm}, \citenamefont {Cava}, \citenamefont {Kivelson}, \citenamefont
  {Nocera}, \citenamefont {Norman},\ and\ \citenamefont
  {Senthil}}]{Broholm2020}%
  \BibitemOpen
  \bibfield  {author} {\bibinfo {author} {\bibfnamefont {C.}~\bibnamefont
  {Broholm}}, \bibinfo {author} {\bibfnamefont {R.~J.}\ \bibnamefont {Cava}},
  \bibinfo {author} {\bibfnamefont {S.~A.}\ \bibnamefont {Kivelson}}, \bibinfo
  {author} {\bibfnamefont {D.~G.}\ \bibnamefont {Nocera}}, \bibinfo {author}
  {\bibfnamefont {M.~R.}\ \bibnamefont {Norman}},\ and\ \bibinfo {author}
  {\bibfnamefont {T.}~\bibnamefont {Senthil}},\ }\bibfield  {title} {\enquote
  {\bibinfo {title} {Quantum spin liquids},}\ }\href
  {https://doi.org/10.1126/science.aay0668} {\bibfield  {journal} {\bibinfo
  {journal} {Science}\ }\textbf {\bibinfo {volume} {367}},\ \bibinfo {pages}
  {eaay0668} (\bibinfo {year} {2020})},\ \Eprint
  {https://arxiv.org/abs/https://www.science.org/doi/pdf/10.1126/science.aay0668}
  {https://www.science.org/doi/pdf/10.1126/science.aay0668} \BibitemShut
  {NoStop}%
\bibitem [{\citenamefont {R\"omling}\ and\ \citenamefont
  {Kamra}(2024)}]{Roemling2024}%
  \BibitemOpen
  \bibfield  {author} {\bibinfo {author} {\bibfnamefont {A.-L.~E.}\
  \bibnamefont {R\"omling}}\ and\ \bibinfo {author} {\bibfnamefont
  {A.}~\bibnamefont {Kamra}},\ }\bibfield  {title} {\enquote {\bibinfo {title}
  {Quantum sensing of antiferromagnetic magnon two-mode squeezed vacuum},}\
  }\href {https://doi.org/10.1103/PhysRevB.109.174410} {\bibfield  {journal}
  {\bibinfo  {journal} {Phys. Rev. B}\ }\textbf {\bibinfo {volume} {109}},\
  \bibinfo {pages} {174410} (\bibinfo {year} {2024})}\BibitemShut {NoStop}%
\bibitem [{\citenamefont {R\"omling}\ \emph {et~al.}(2025)\citenamefont
  {R\"omling}, \citenamefont {Feist}, \citenamefont {Garc\'{\i}a-Vidal},\ and\
  \citenamefont {Kamra}}]{Roemling2025}%
  \BibitemOpen
  \bibfield  {author} {\bibinfo {author} {\bibfnamefont {A.-L.~E.}\
  \bibnamefont {R\"omling}}, \bibinfo {author} {\bibfnamefont {J.}~\bibnamefont
  {Feist}}, \bibinfo {author} {\bibfnamefont {F.~J.}\ \bibnamefont
  {Garc\'{\i}a-Vidal}},\ and\ \bibinfo {author} {\bibfnamefont
  {A.}~\bibnamefont {Kamra}},\ }\bibfield  {title} {\enquote {\bibinfo {title}
  {Squeezing and quantum control of the antiferromagnetic magnon pseudospin},}\
  }\href {https://doi.org/10.1103/qjm2-d19g} {\bibfield  {journal} {\bibinfo
  {journal} {Phys. Rev. B}\ }\textbf {\bibinfo {volume} {112}},\ \bibinfo
  {pages} {214444} (\bibinfo {year} {2025})}\BibitemShut {NoStop}%
\bibitem [{\citenamefont {Skogvoll}\ \emph {et~al.}(2021)\citenamefont
  {Skogvoll}, \citenamefont {Lidal}, \citenamefont {Danon},\ and\ \citenamefont
  {Kamra}}]{Skogvoll2021}%
  \BibitemOpen
  \bibfield  {author} {\bibinfo {author} {\bibfnamefont {I.~C.}\ \bibnamefont
  {Skogvoll}}, \bibinfo {author} {\bibfnamefont {J.}~\bibnamefont {Lidal}},
  \bibinfo {author} {\bibfnamefont {J.}~\bibnamefont {Danon}},\ and\ \bibinfo
  {author} {\bibfnamefont {A.}~\bibnamefont {Kamra}},\ }\bibfield  {title}
  {\enquote {\bibinfo {title} {Tunable anisotropic quantum rabi model via a
  magnon--spin-qubit ensemble},}\ }\href
  {https://doi.org/10.1103/PhysRevApplied.16.064008} {\bibfield  {journal}
  {\bibinfo  {journal} {Phys. Rev. Appl.}\ }\textbf {\bibinfo {volume} {16}},\
  \bibinfo {pages} {064008} (\bibinfo {year} {2021})}\BibitemShut {NoStop}%
\bibitem [{\citenamefont {Kittel}(1953)}]{Kittel1953}%
  \BibitemOpen
  \bibfield  {author} {\bibinfo {author} {\bibfnamefont {C.}~\bibnamefont
  {Kittel}},\ }\href@noop {} {\emph {\bibinfo {title} {Introduction to Solid
  State Physics}}}\ (\bibinfo  {publisher} {John Wiley \& Sons},\ \bibinfo
  {address} {New York},\ \bibinfo {year} {1953})\BibitemShut {NoStop}%
\bibitem [{\citenamefont {Satyanarayana}(1985)}]{Satyanarayana1985}%
  \BibitemOpen
  \bibfield  {author} {\bibinfo {author} {\bibfnamefont {M.~V.}\ \bibnamefont
  {Satyanarayana}},\ }\bibfield  {title} {\enquote {\bibinfo {title}
  {Generalized coherent states and generalized squeezed coherent states},}\
  }\href {https://doi.org/10.1103/PhysRevD.32.400} {\bibfield  {journal}
  {\bibinfo  {journal} {Phys. Rev. D}\ }\textbf {\bibinfo {volume} {32}},\
  \bibinfo {pages} {400--404} (\bibinfo {year} {1985})}\BibitemShut {NoStop}%
\bibitem [{\citenamefont {Kim}, \citenamefont {de~Oliveira},\ and\
  \citenamefont {Knight}(1989)}]{Kim1989}%
  \BibitemOpen
  \bibfield  {author} {\bibinfo {author} {\bibfnamefont {M.~S.}\ \bibnamefont
  {Kim}}, \bibinfo {author} {\bibfnamefont {F.~A.~M.}\ \bibnamefont
  {de~Oliveira}},\ and\ \bibinfo {author} {\bibfnamefont {P.~L.}\ \bibnamefont
  {Knight}},\ }\bibfield  {title} {\enquote {\bibinfo {title} {Properties of
  squeezed number states and squeezed thermal states},}\ }\href
  {https://doi.org/10.1103/PhysRevA.40.2494} {\bibfield  {journal} {\bibinfo
  {journal} {Phys. Rev. A}\ }\textbf {\bibinfo {volume} {40}},\ \bibinfo
  {pages} {2494--2503} (\bibinfo {year} {1989})}\BibitemShut {NoStop}%
\bibitem [{\citenamefont {Walls}\ and\ \citenamefont
  {Milburn}(2025)}]{Walls2025QuantumOptics}%
  \BibitemOpen
  \bibfield  {author} {\bibinfo {author} {\bibfnamefont {D.~F.}\ \bibnamefont
  {Walls}}\ and\ \bibinfo {author} {\bibfnamefont {G.~J.}\ \bibnamefont
  {Milburn}},\ }\href {https://doi.org/10.1007/978-3-031-84177-4} {\emph
  {\bibinfo {title} {Quantum Optics}}},\ \bibinfo {edition} {3rd}\ ed.,\
  Graduate Texts in Physics\ (\bibinfo  {publisher} {Springer},\ \bibinfo
  {address} {Cham},\ \bibinfo {year} {2025})\BibitemShut {NoStop}%
\bibitem [{\citenamefont {Breuer}\ and\ \citenamefont
  {Petruccione}(2010)}]{Breuer}%
  \BibitemOpen
  \bibfield  {author} {\bibinfo {author} {\bibfnamefont {H.-P.}\ \bibnamefont
  {Breuer}}\ and\ \bibinfo {author} {\bibfnamefont {F.}~\bibnamefont
  {Petruccione}},\ }\bibfield  {title} {\enquote {\bibinfo {title} {The theory
  of open quantum systems},}\ }\href
  {https://doi.org/10.1093/acprof:oso/9780199213900.001.0001} {\bibfield
  {journal} {\bibinfo  {journal} {Oxford Academic}\ } (\bibinfo {year}
  {2010})}\BibitemShut {NoStop}%
\bibitem [{\citenamefont {Xu}\ \emph {et~al.}(2023{\natexlab{b}})\citenamefont
  {Xu}, \citenamefont {Gu}, \citenamefont {Li}, \citenamefont {Weng},
  \citenamefont {Wang}, \citenamefont {Li}, \citenamefont {Wang}, \citenamefont
  {Zhu},\ and\ \citenamefont {You}}]{Xu2023_magnon}%
  \BibitemOpen
  \bibfield  {author} {\bibinfo {author} {\bibfnamefont {D.}~\bibnamefont
  {Xu}}, \bibinfo {author} {\bibfnamefont {X.-K.}\ \bibnamefont {Gu}}, \bibinfo
  {author} {\bibfnamefont {H.-K.}\ \bibnamefont {Li}}, \bibinfo {author}
  {\bibfnamefont {Y.-C.}\ \bibnamefont {Weng}}, \bibinfo {author}
  {\bibfnamefont {Y.-P.}\ \bibnamefont {Wang}}, \bibinfo {author}
  {\bibfnamefont {J.}~\bibnamefont {Li}}, \bibinfo {author} {\bibfnamefont
  {H.}~\bibnamefont {Wang}}, \bibinfo {author} {\bibfnamefont {S.-Y.}\
  \bibnamefont {Zhu}},\ and\ \bibinfo {author} {\bibfnamefont {J.~Q.}\
  \bibnamefont {You}},\ }\bibfield  {title} {\enquote {\bibinfo {title}
  {Quantum control of a single magnon in a macroscopic spin system},}\ }\href
  {https://doi.org/10.1103/PhysRevLett.130.193603} {\bibfield  {journal}
  {\bibinfo  {journal} {Phys. Rev. Lett.}\ }\textbf {\bibinfo {volume} {130}},\
  \bibinfo {pages} {193603} (\bibinfo {year} {2023}{\natexlab{b}})}\BibitemShut
  {NoStop}%
\bibitem [{\citenamefont {Dey}\ \emph {et~al.}()\citenamefont {Dey},
  \citenamefont {Verma}, \citenamefont {Weiler},\ and\ \citenamefont
  {Kamra}}]{Dey2025}%
  \BibitemOpen
  \bibfield  {author} {\bibinfo {author} {\bibfnamefont {B.}~\bibnamefont
  {Dey}}, \bibinfo {author} {\bibfnamefont {S.}~\bibnamefont {Verma}}, \bibinfo
  {author} {\bibfnamefont {M.}~\bibnamefont {Weiler}},\ and\ \bibinfo {author}
  {\bibfnamefont {A.}~\bibnamefont {Kamra}},\ }\href
  {https://doi.org/https://doi.org/10.48550/arXiv.2507.19066} {\enquote
  {\bibinfo {title} {Sensing magnonic quantum superpositions using a bosonic
  mode as the probe},}\ }\Eprint {https://arxiv.org/abs/arXiv.2507.19066}
  {arXiv.2507.19066} \BibitemShut {NoStop}%
\bibitem [{Tar()}]{Tarek}%
  \BibitemOpen
  \href@noop {} {}\bibinfo {note} {T.~Moussa et al., Lindblad description of
  dispersively coupled quantum systems, (unpublished).}\BibitemShut {Stop}%
\end{thebibliography}%

\end{document}